\RequirePackage{fix-cm}
 \documentclass{svjour3}   
 \usepackage[right=1.0in, left=1.0in, top=1.2in, bottom=1.0in]{geometry}
\smartqed  
\usepackage{graphicx}
\usepackage{epstopdf, epsfig}
\usepackage{color}
\usepackage{float}
\usepackage{amsmath}
\usepackage{amssymb}
\usepackage{amsfonts}
\usepackage{tabularx}
\usepackage{bm}
\usepackage{natbib}
\usepackage{bm}
\usepackage{array}
\usepackage{tabularx}
\usepackage{multirow,multicol, makecell}
\usepackage{float}
\usepackage{breqn}
\usepackage{mathtools}
\usepackage{breqn}
\usepackage[flushleft]{threeparttable}
\usepackage{booktabs,caption}
\usepackage{footnote}
\usepackage{comment}
\makesavenoteenv{tabular}
\newcolumntype{C}[1]{>{\centering\arraybackslash}p{#1}}
\newcolumntype{L}[1]{>{\raggedright\arraybackslash}p{#1}}
\newcolumntype{M}[1]{>{\centering\arraybackslash}m{#1}}
 \usepackage[mathlines,displaymath]{lineno}

\begin{document}

\title{Error propagation dynamics of PIV-based pressure field calculation~(3): What is the minimum resolvable pressure in a reconstructed field?
}

\titlerunning{PIV-based pressure: minimum resolvable pressure}        

\author{Mingyuan Nie       \and
        Jared P. Whitehead \and
        Geordie Richards   \and
        Barton L. Smith   \and
        Zhao Pan 
}

\authorrunning{M. Nie, J. P. Whitehead, G. Richards, B. L. Smith, Z. Pan,} 

\institute{M. Nie \at
              Mechanical and Mechatronics Engineering  \\
              University of Waterloo\\
              Waterloo, ON N2L 3G1, Canada\\
              \email{mingyuan.nie@uwaterloo.ca}           
           \and
        	J. P. Whitehead \at
			Mathematics Department\\
			Brigham Young University\\
			Provo, UT 84602, USA\\
			\email{whitehead@mathematics.byu.edu}
		   \and
			G. Richards \at
			 Department of Mathematical and Computational Sciences \\
			University of Toronto Mississauga\\
			Mississauga, ON, L5L 1C6, Canada\\
			\email{ geordie.richards@utoronto.ca}
           \and 
            B. L. Smith\at
            Mechanical and Aerospace Engineering \\
            Utah State University\\
            Logan, UT, 84322, USA\\
            \email{barton.smith@usu.edu}
            \and 
            Z. Pan\at
            Mechanical and Mechatronics Engineering  \\
              University of Waterloo\\
              Waterloo, ON N2L 3G1, Canada\\
              \email{zhao.pan@uwaterloo.ca}   
}


\maketitle

\begin{abstract}
An analytical framework for the propagation of velocity errors into PIV-based pressure calculation is extended. 
Based on this framework, the optimal spatial resolution and the corresponding minimum field-wide error level in the calculated pressure field are determined. 
This minimum error can be viewed as the smallest resolvable pressure. 
We find that the optimal spatial resolution is a function of the flow features (patterns and length scales), fundamental properties of the flow domain (e.g., geometry of the flow domain and the type of the boundary conditions), in addition to the error in the PIV experiments, and the choice of numerical methods.
Making a general statement about pressure sensitivity is difficult. 
The minimum resolvable pressure depends on competing effects from the experimental error due to PIV and the truncation error from the numerical solver, which is affected by the formulation of the solver.
This means that PIV experiments motivated by pressure measurements should be carefully designed so that the optimal resolution (or close to the optimal resolution) is used.
Flows (Re=$1.27 \times 10^4$ and $5\times 10^4$) with exact solutions are used as examples to validate the theoretical predictions of the optimal spatial resolutions and pressure sensitivity. 
The numerical experimental results agree well with the rigorous analytical predictions.
We also propose an a \textit{posterior} method to estimate the contribution of truncation error using Richardson extrapolation and that of PIV error by adding artificially overwhelming noise.
We also provide an introductory analysis of the effects of interrogation window overlap in PIV in the context of the pressure calculation. 
\end{abstract}

\section{Introduction}
\label{sec:Intro}
\subsection{Background and Motivation}
\label{sec:Intro:Background}

Experimental pressure measurements are useful for determining loads on structures and for examination of acoustic effects. 
However, it has not been possible until recently to determine the pressure field away from surfaces. 
In the last decade, it has been shown that the pressure field may be determined from velocity data measured using Particle Image Velocimetry (PIV). 
PIV-based pressure calculation techniques have received significant  interest because the output provides field measurements with high frequency response \citep{van2013piv}. 
The accuracy of pressure fields derived from PIV data has improved with the PIV technique itself, as hardware improvements provide increased spatial and temporal resolution and new techniques have provided a fully three-dimensional velocity field \citep{scarano2012tomographic,hinsch2002holographic}. 

With these recent advances, a natural inquiry is ``How accurate is the pressure field obtained from PIV?'' or ``How does accuracy in the velocity field affect the accuracy of the computed pressure field?'' or ``What is the minimum measurable pressure?'' This paper will address these questions, but first we survey the state-of-the-art in PIV-based pressure field calculations.

Velocimetry-based pressure reconstruction is a straight-forward idea that can be traced back to \cite{schwabe1935druckermittlung}. Technical limitations of the imaging technique in \cite{schwabe1935druckermittlung} (i.e., low spatial and temporal resolution, etc.) and consequently large error in the velocity field measurement, led to a calculated pressure that was not reliable enough to ensure any quantitative confidence at the time. After more than 25 years of development, PIV has become a standard and reliable non-invasive velocity field measurement technique \cite{adrian2005twenty}, which not only provides the vector velocity field measurements but also describes the uncertainty of these measurements \citep{timmins2012method, sciacchitano2013piv,charonko2013estimation,wieneke2015piv}. 

Built on these advancements in PIV techniques, calculation of the PIV-based pressure field has become common. PIV-based pressure field reconstruction methods are divided into two categories: i) directly integrating the pressure gradient from the Navier-Stokes equations, and ii) integrating the Poisson equation derived from the Navier-Stokes equation to obtain the scalar pressure field. Direct integral methods usually involve multi-path integral techniques that improve robustness of the algorithm. Typical examples include four-path integral \citep{baur1999piv}, and the omni-directional integral method \citep{liu2006instantaneous, wang2019gpu}, which requires $2M(N+M) + 2N(2M+N)$ integral paths for an $M\times N$ mesh. 

Recent algorithmic considerations investigate alternative numerical methods such as a least squares linear system solver \citep{jeon2015least}, and spectral decomposition \citep{wang2017spectral} to achieve a robust and fast pressure solution by numerically integrating the pressure gradient. Pressure-Poisson-equation-based methods often involve well-defined explicit boundary conditions such as Neumann, Dirichlet or, more often, mixed boundary conditions, which have a straightforward physical and mathematical interpretation. Examples may be found in \citet{de2012instantaneous}, \citet{probsting2013estimation}, and \cite{de2013pressure}. In recent years, it is popular to use data assimilation or machine learning to solve the necessary partial differential equations. Examples can be found in \citet{lagaris1998artifical, sirignano2018dgm, li2021deep}, where artificial neural networks are involved to solve the equations. For either the pressure gradient integration or the Poisson equation approach, the state-of-the-art implementation uses time-resolved 2D and/or 3D PIV data with numerically optimized solvers.

{Taking advantage of the advancing velocimetry techniques and velocimetry-based pressure reconstruction algorithms, PIV-based pressure reconstruction techniques have been increasingly applied to various fields of study.} Examples in classic topics include pressure field and loads on airfoils \citep{violato2011lagrangian, jeon20163d}, wind turbine blades \citep{lignarolo2014experimental,villegas2014evaluation}, water slamming of a wedge \citep{panciroli2013evaluation} and a boat hull \citep{porfiri2017new}, pressure distribution in turbulent boundary layers \citep{ghaemi2012piv,zhang2017deformation}, as well as instantaneous pressures calculation by using Taylor's frozen turbulence hypothesis \citep{van2019pressure}. PIV-based pressure calculations are also useful for bio-fluidic studies such as the pressure field in a glottal channel \citep{oren2015intraglottal}. The extended applications cover aero-acoustics with acoustic analogies \citep{haigermoser2009application,koschatzky2011study,moore2011two,nickels2017acoustic,leon2017measurement}, and compressible flows \cite{van2007evaluation,van2008principles}.

Fundamental research on PIV-based pressure calculations has led to novel algorithm development and optimization for specific applications. For example, \cite{charonko2010assessment} benchmarked several different pressure field reconstruction algorithms and found that the performance of the PIV-based pressure calculation is affected by almost every factor involved in the experiments (e.g., type of flow, spatial and temporal resolutions, filtering of the PIV data, the type of numerical solver, and error level in the PIV data). There is no universal optimal experimental setup for all applications, although, as shown below for a specific problem, optimal spatial and temporal resolutions do exist that minimize the error in the calculated pressure field. \cite{de2012instantaneous} pointed out that a numerical Poisson solver acts as a low-pass filter, which tends to eliminate the high frequency information from the PIV experiments.

In subsequent work, \cite{faiella2021error} showed that this low-pass filter is not due to the specific numerical scheme, but is rooted in the properties of the Poisson operator. Thus, low frequency error due to PIV measurements should be avoided to minimize the error that propagates to the calculated pressure. A more general study of error propagation of the PIV-based pressure field calculation showed that the geometry (dimension, shape, and size) of the domain and type of boundary conditions impact the error propagation as well \citep{pan2016error1}. It was also shown that pure Neumann boundary conditions should be avoided. 
In \citep{van2017comparative} a new series of benchmarks were performed through the NIOPLEX project. 
In this benchmark, a variety of pressure field reconstruction methods (velocity field from PIV and Lagrangian particle tracking (LPT)) were applied to a high-speed subsonic compressible flow over an axisymmetric step. 
They found that noise in the velocimetry measurements reduced the accuracy of the pressure reconstruction, yet LPT-based techniques produce more accurate pressure fields than the PIV-based approaches. 

New pressure field reconstruction methods developed for unstructured data collected by high seeding density LPT are presumably compatible with volumetric PIV data, which are on structured mesh. 
For example, \citet{bobrov2021pressure} applied weighted least-squares to the Navier Stokes and/or Poisson equations, in which the weighting coefficients  correspond to the distances between particles.
Through validation based on DNS, the algorithm is shown to be robust and accurate compared to simulated unstructured LPT data in 3D.
\citet{sperotto2021meshless} use two layers of least-squares and Radial Basis Functions (RBFs) to approximate the Lagrangian velocity field and then solve for the pressure field by integrating a pressure Poisson equation. 
However, these two example works did not thoroughly investigate the uncertainty quantification of these methods.

Despite numerous recent studies on PIV-based pressure calculations as a quantitative measurement technique, the uncertainty of the technique and how it depends on velocity accuracy has not been sufficiently addressed. Only a few works have covered this topic. \cite{azijli2016posteriori} proposed a posteriori uncertainty quantification method of PIV-based pressure calculations under a Bayesian framework. To the best of our knowledge, this research was the first work that provided direct quantification of the uncertainty in the reconstructed pressure field. In addition to requiring several simplifying assumptions, the approach taken in \cite{azijli2016posteriori} will not provide error estimation \textit{a priori}, and little analytical insight to the nature of the error propagation from the velocity field to the pressure field is provided. \cite{pan2016error1} proposed an upper bound on the error in the calculated pressure field which is a function of the fundamental factors of the flow field, such as geometry of the domain and the type of boundary conditions. Even though the upper bound is not always apparent and could overestimate the error, it can be considered an \textit{a priori} estimate of the worst possible error level in the reconstructed pressure field and thus aid the experimental design and optimization. 
More recently, \citet{mcclure2017optimization} proposed an estimation of the optimal spatial and temporal resolution that minimize the error in the pressure field, but did not provide the minimum error in the reconstructed pressure, which can be interpreted as the sensitivity of the pressure reconstruction. 
 \citet{zhang2022uncertainty} proposed a straightforward practical strategy that models the pressure propagation from PIV data to the pressure field as a linear transformation, which is rooted in the specific numerical solver that is used for pressure reconstruction. 
Using various pressure solvers as validation examples, this method has been shown to be able to predict the covariance of the error in the pressure field given that the \textit{a priori} uncertainty quantification of the PIV data is accurate, and the exact formulation of the pressure solver (including grid spacing ) is known.
Despite the practicality and effectiveness of this approach, it is a \textit{posterior} uncertainty estimation similar to \cite{azijli2015solenoidal} but demands much lower computational cost. In addition, this method does not provide analytical insight into the design of experiments (e.g. the choice of resolution of PIV) and the explicit mechanism of how the error in the velocity field propagates to the pressure field.

In the current study, we will begin to answer one of the fundamental questions posed above: ``What is the minimum resolution or the sensitivity of the PIV-based pressure calculation for a given experimental setup, and what is the optimal spatial resolution for a PIV experiment with pressure reconstruction being the end goal?'' 

\subsection{The focus of the current work in a bigger picture}
\label{sec:Intro:focus}
The brief summary above shows that velocimetry-based pressure field reconstruction and its uncertainty quantification is a complex problem for which every aspect in computation and data acquisition matters. 
Even some subtle factors can significantly impact performance of the pressure reconstruction.

\begin{figure}[!h]
	\centering
	\includegraphics[width=1\textwidth]{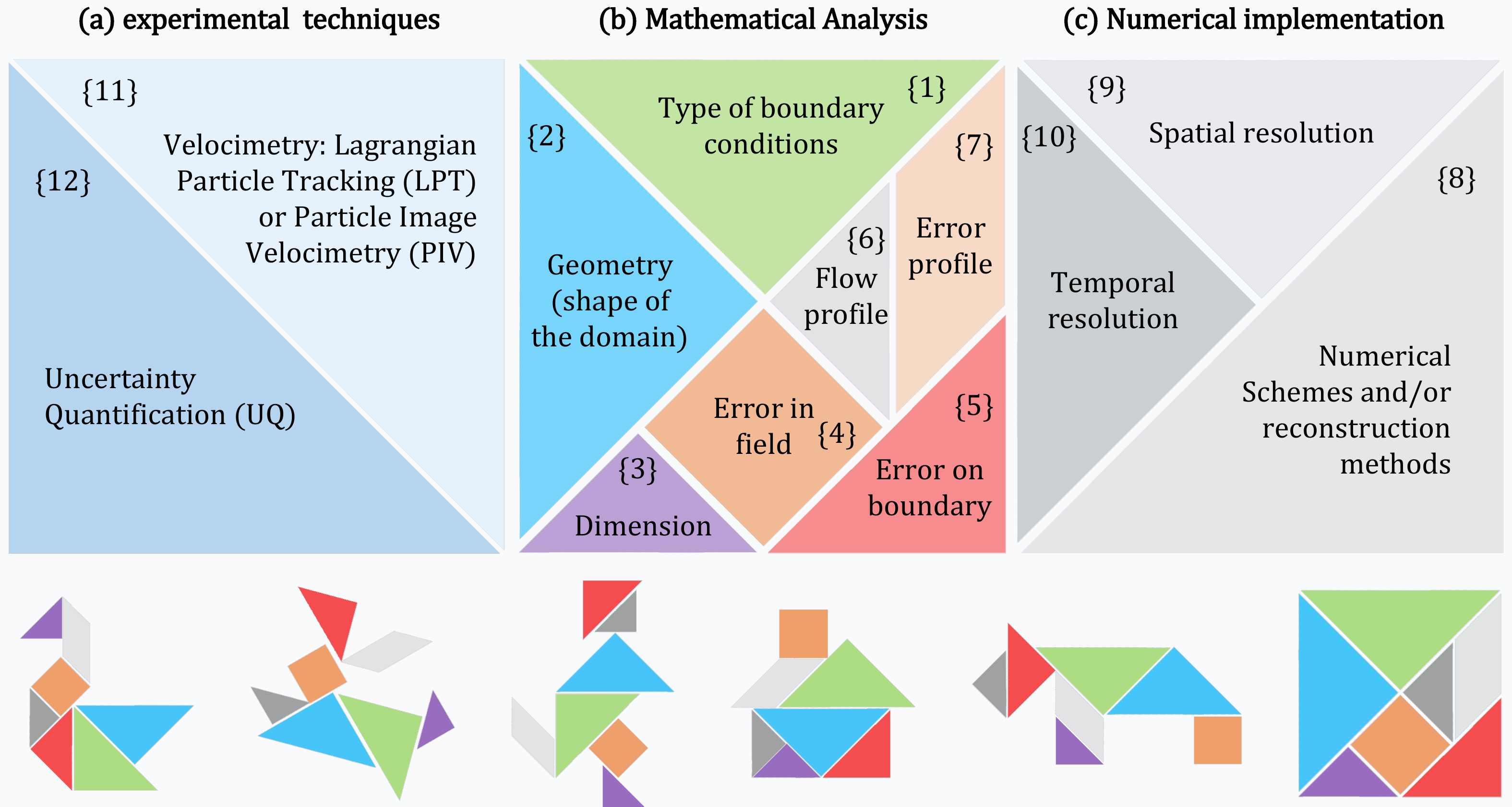}
	\caption{Error propagation of the PIV-based pressure reconstruction can be thought of as three Tangrams. A Tangram is a geometric Chinese puzzle that can be rearranged to make various shapes. Here, Puzzle~(a) represents various velocimetry techniques and the corresponding uncertainty quantification. Tangram~(b) represents the mathematical analysis of the error problem, focusing on the details of the mathematical construction of the pressure field. Puzzle~(c) represents the numerical issues that arise from the pressure reconstruction. The lower row shows how the Tangrams are traditionally played to form different shapes. The authors want to convey a view that ``no matter how to play the game, you always play with some or all of these puzzle pieces.''
		\label{fig:puzzle}}
\end{figure}

We propose that the dynamics of the error propagation can be analogized in a seven-piece Tangram puzzle shown in Fig.~\ref{fig:puzzle} where each piece of the puzzle represents one important factor.
As an example, the size and shape of the domain (see Fig.~\ref{fig:puzzle}(b)) can affect the error propagation dramatically. 
The impact of the domain geometry is inherently rooted in the properties of the Laplace operator, which plays a central role when the pressure field is reconstructed based on the pressure Poisson equation. The colored shapes in this big picture (Fig~\ref{fig:puzzle}) represent areas that are well-studied and/or have recently been explained, whereas the grey areas are those that are largely unexplored (different shades of gray represents the different difficulty of the each sub-problem in the authors' point of view: the darker the more difficult). 
For example, pieces~\{1\}~--~\{5\} have been experimentally reported by \cite{charonko2010assessment} and analytically addressed in \cite{pan2016error1}. 
The impact of pieces~\{6\} and \{7\} was first observed by \cite{charonko2010assessment,de2013pressure}, respectively, and later partially covered in \cite{faiella2021error} analytically.

In practice, the numerical implementation of the pressure solver also involves three important factors.
Two of these are spatial resolution and temporal resolution.
The third is the numerical schemes/methods for the pressure field reconstruction~(Fig.~\ref{fig:puzzle}(c)). In addition to many studies involving different numerical schemes (e.g. different benchmarking pressure solvers employed in \cite{charonko2010assessment}, and recent novel solvers in \cite{jeon2015least,wang2017spectral,sperotto2021meshless}) that could provide empirical insights into pieces~\{8\}~--~\{10\}, a work directly addressing the impact of both spatial and temporal resolutions can be found in \cite{mcclure2017optimization}.

Moreover, the velocity measurement technique (e.g., PIV or LPT) and the corresponding errors  are also two fundamental aspects of the problem. Before discussing error propagation, understanding the source of the error is foundational. 
In the context of the velocimetry-based pressure field reconstruction, the performance and features of measurement techniques, such as planar or volumetric, Lagrangian Particle Tracking (LPT) or PIV (puzzle piece~\{11\}, review can be found in \citet{raffel2018particle, sciacchitano2019uncertainty}) as well as the corresponding uncertainty quantification (puzzle piece~\{12\}, see references \citet{smith2001measuring, sciacchitano2016piv,rajendran2020uncertainty}) are important. 
High accuracy velocimetry with reliable uncertainty quantification perhaps is the ultimate solution to high accuracy pressure-field reconstruction.

These previous multi-party efforts as well as connected or disconnected understandings of the complete picture of the error propagation dynamics have laid the groundwork for exploring the rest of the puzzle analytically. 
In the current work, we focus on a preliminary semi-analytical investigation on the impact of piece~\{9\} (spatial resolution of PIV experiments) based on recent understanding of pieces~\{1\}~--~\{5\}, and \{7\}. 
In the next section we develop a basic theory to explain these effects. 
Numerical experiments are then used to validate the theoretical predictions.


\section{Problem setup and definitions}
\label{sec:Problem}
PIV-based pressure calculation is rooted in the Navier-Stokes equations. {Arranging the nondimensionalized Navier-Stokes equations we have} 
\begin{linenomath}
	\begin{equation}
	\label{eq:NSeq}
	\nabla p = -\left( \frac{\partial \bm{u}}{\partial t} + \left(\bm{u}\cdot\nabla\right) \bm{u} - \frac{1}{Re} \nabla^2 \bm{u}  \right),
	\end{equation}
\end{linenomath}
where $\bm{u}$ is the velocity field, which is obtained from experiments, and $p$ is the pressure field, which is to be determined. $Re$ is the Reynolds number. 

As noted above, PIV-based pressure field calculation methods fall into two categories: i) direct integration of the pressure gradient ($\nabla p$) (e.g.,\cite{schwabe1935druckermittlung,liu2006instantaneous}) from eq.~\eqref{eq:NSeq}, ii) applying the divergence operator to eq.~\eqref{eq:NSeq} and solving the corresponding Poisson equation with respect to the pressure field $p$: 
\begin{linenomath}
	\begin{equation}
	\label{eq:Poisson1}
	\nabla^2 p = f(\bm{u}) = -\nabla \cdot\left( \frac{\partial \bm{u}}{\partial t} + \left(\bm{u}\cdot\nabla\right) \bm{u} - \frac{1}{Re} \nabla^2 \bm{u}  \right),
	\end{equation}
\end{linenomath}
where the right hand side $f(\bm{u})$ is called the ``data'' \citep{pan2016error1}.\footnote{We will adopt this terminology in the current paper due to the nature of this study (e.g., \cite{fraenkel2000introduction}.) To prevent any confusion, we will address the experimental data from PIV as ``experimental results'' or ``PIV results''.}
In this study, we focus on the latter method.
Note that in the following discussions, the time derivative term $\frac{\partial \bm{u}}{\partial t}$ vanishes for an incompressible flow as $\nabla \cdot \frac{\partial \bm{u}}{\partial t} = 0$. 
When $Re$ is large, the viscous term can further be neglected \citep{de2012instantaneous, van2013piv} and eq.~\eqref{eq:Poisson1} can be simplified as:
\begin{linenomath}
	\begin{align}
		\label{eq:PoissonEq}
		\nabla^2 p &= f(\bm{u}) = - \nabla \cdot \left( \left( \bm{u} \cdot \nabla\right) \bm{u} \right) && \quad \text{in}  ~\Omega,	
	\end{align}
\end{linenomath}
where $p$ and $\bm{u}$ are the pressure and velocity field respectively. 
The two-dimensional form of eq.~\eqref{eq:PoissonEq}, for example, is 
\begin{linenomath}
	\begin{align}
\label{eq:PoissonEq2D}
	\nabla^2 p = - \left(\left(\frac{\partial u}{\partial x}\right)^2\ + 2\left(\frac{\partial u}{\partial y}\right) \cdot \left(\frac{\partial v}{\partial x}\right) + \left(\frac{\partial v}{\partial y}\right)^2\right) && \quad \text{in}  ~\Omega \in \mathbb{R}^2,  
	\end{align}
\end{linenomath}
and is widely used in velocimetry-based pressure reconstruction literature \citep{de2012instantaneous, sperotto2021meshless}.

Eq.~\eqref{eq:Poisson1} must be solved with proper boundary conditions (BCs) such as Dirichlet (enforced pressure on the boundary), and/or Neumann (enforced pressure gradient on the boundary) boundary conditions. Thus a complete description of the problem in the domain $\Omega$ can be described as
\begin{linenomath}
	\begin{equation}
	\label{eq:PDE}
	\nabla^2 p = f(\bm{u}) \qquad \text{in} ~ \Omega, 
	\end{equation}
\end{linenomath}
with Neumann BCs,
\begin{linenomath}
	\begin{equation}
	\label{eq:NBC}
	\nabla p \cdot \mathbf{n} = g(\bm{u}) \quad \text{on} ~ \partial\Omega, 
	\end{equation}
\end{linenomath}
and/or Dirichlet BCs
\begin{linenomath}
	\begin{equation}
	\label{eq:DBC}
	p = h(\bm{u}) \qquad \text{on} ~ \partial\Omega, 
	\end{equation}
\end{linenomath}
where $f(\bm{u})$, $g(\bm{u})$ and $h(\bm{u})$ are corresponding functions of the velocity field. $g(\bm{u})$ often takes a form similar to eq.~\eqref{eq:NSeq} which can be directly evaluated from the velocity field measured on the boundary: 
\begin{linenomath}
	\begin{equation}
	\nabla p \cdot \mathbf{n}= -\left( \frac{\partial \bm{u}}{\partial t} + \left(\bm{u}\cdot\nabla\right) \bm{u} - \frac{1}{Re} \nabla^2 \bm{u}  \right) \cdot\mathbf{n},
	\end{equation}
\end{linenomath}
and the Dirichlet BCs can be either directly measured from pressure transducers or calculated from the Bernoulli equations (e.g.,~\cite{de2012instantaneous}).

Clearly, the error or noise, from experimental measurements will propagate to the calculated pressure field. 
However, the typical propagation analysis using Taylor Series Method or Monte Carlo methods \citep{coleman2009experimentation} are difficult since the experimental data reduction equation becomes even more complex than eq.~\eqref{eq:Poisson1}. In the current study, we directly analyze the error propagation of the PIV-based pressure calculation using the underlying equations eq.~\eqref{eq:PDE} and corresponding boundary conditions. 
Denoting the error in the measured velocity field as $\bm{\epsilon}_u$ and the true value of the velocity field as $\bm{u}$, the error contaminated velocity measurement $\tilde{\bm{u}}$ can be modeled as $\tilde{\bm{u}} =\bm{u} + \bm{\epsilon}_u $. 
Similarly, the noisy calculated pressure field $\tilde{p}$ can be modeled as $\tilde{p} = p + \epsilon_p$, where $\epsilon_p$ is the error in the calculated pressure field and $p$ is the unknown true value. 
With measurement error considered, $\eqref{eq:PDE}$, $\eqref{eq:NBC}$, and $\eqref{eq:DBC}$ are implemented in practice as 
\begin{linenomath}
	\begin{equation}
	\label{eq:PDE1}
	\nabla^2 \tilde{p} = f(\tilde{\bm{u}}) \qquad \text{in} ~ \Omega, 
	\end{equation}
\end{linenomath}
with Neumann BCs,
\begin{linenomath}
	\begin{equation}
	\label{eq:NBC1}
	\nabla \tilde{p} \cdot \mathbf{n} = g(\tilde{\bm{u}}) \qquad \text{on} ~ \partial\Omega, 
	\end{equation}
\end{linenomath}
and/or Dirichlet BCs
\begin{linenomath}
	\begin{equation}
	\label{eq:DBC1}
	\tilde{p} = h(\tilde{\bm{u}}) \qquad \text{on} ~ \partial\Omega.
	\end{equation}
\end{linenomath}
We will further quantify the relationship between $\bm{\epsilon}_u$ and $\epsilon_p$.
To adequately perform this comparison, we use the space-averaged $L^2$-norm of a field on the domain $\Omega$ defined as
\begin{linenomath}
	\begin{equation}
	\label{eq:L2norm}
	||\epsilon||_{L^2(\Omega)} = \sqrt{\frac{\int \epsilon^2 d\Omega}{|\Omega|}}, 
	\end{equation}
\end{linenomath}
where $|\Omega|$ denotes the area or volume of the domain depending on the dimensions of the domain. 
This choice of the measure is beneficial in three ways: i) it has a straightforward physical meaning of the ``power'' of error per unit space; ii) it makes the later mathematical analysis tractable; and iii) the discrete form of eq.~\eqref{eq:L2norm} is a root mean square (RMS) measure of error which is often considered as an ``effective value'' of a given dependent variable . 
In the current work, we call this space-averaged error ($||\epsilon||_{L^2(\Omega)}$) the `error level'. 
The relationship between $\bm{\epsilon}_u$ and $\epsilon_p$ is an analysis of the propagation from the error level in the velocity field ($||\bm{\epsilon}_u||_{L^2(\Omega)}$) to the error level in the pressure field ($||\epsilon_p||_{L^2(\Omega)}$).

\section{Error estimation of reconstructed pressure field}
\label{sec:Derivation}

\subsection{Theory and the physical interpretation}
\label{sec:Derivation:Theory}
For the purposes of this study, we consider a two dimensional flow on a structured mesh with grid spacing $h\times h$. 
For simplicity, in the following sections we assume that the measured velocity field from the PIV experiments has point-wise independent \footnote{This assumption is not necessary, which will be discussed in section~\ref{sec:Practical:Overlap}. Making this assumption is solely to simplify presentation of the core idea.}, zero-mean  noise with variance $\sigma_u$ and $\sigma_v$ in the two cardinal directions. 
The expected error level in the calculated pressure field can be estimated as 
\begin{linenomath}
	\begin{equation}
	\label{eq:error}
	\begin{split}
	\|\epsilon_p\|_{(L^2(\Omega))}  &\lesssim \|\epsilon_{p,T} \|_{L^2(\Omega)} + \|\epsilon_{p,E} \|_{L^2(\Omega)} \\
	& \approx \underbrace{C_1 \left(  \left\| \frac{\partial^2 p}{\partial x^2} \right\|_{L^2(\Omega)} + 2\left\| \nabla^{-2}\frac{\partial^4 p}{\partial^2 x \partial^2 y} \right\|_{L^2(\Omega)} +\left\| \frac{\partial^2 p}{\partial x^2} \right\|_{L^2(\Omega)} \right)h^m}_{\text{Truncation~error~contribution}} + \overbrace{C_0 C_2 \left(\frac{\sigma_u^2 + \sigma_v^2}{2}\right) h^n}^{\text{PIV~error~contribution}},
	\end{split}
	\end{equation}
\end{linenomath}
where $||\epsilon_{p,T} ||_{L^2(\Omega)} $ is the truncation error of the numerical scheme arising from the Poisson solver, and the second term ($||\epsilon_{p,E}||_{L^2(\Omega)}$) includes the effect of the experimental errors in the measured velocity field (the derivation of this inequality with greater details can be found in Appendix~\ref{sec:Appendix derivation}). For a specific example, a flow in an $L\times L$ square domain, with pure Dirichlet boundary conditions, the pressure field is solved by a second order Poisson solver with central difference scheme. 
In this setting, eq.~\eqref{eq:error} leads to a more particular form with specific parameters:
\begin{linenomath}
	\begin{equation}
	\label{eq:error1}
	\|\epsilon_p\|_{(L^2(\Omega))} \lesssim \frac{1}{12} \left(  \left\| \frac{\partial^2 p}{\partial x^2} \right\|_{L^2(\Omega)} + 2\left\| \nabla^{-2}\frac{\partial^4 p}{\partial^2 x \partial^2 y} \right\|_{L^2(\Omega)} +\left\| \frac{\partial^2 p}{\partial x^2} \right\|_{L^2(\Omega)} \right)h^2 + 0.901^2 \frac{L^2}{2\pi^2}  \left(\frac{\sigma_u^2 + \sigma_v^2}{2}\right) h^{-2}. 
	\end{equation}
\end{linenomath}
The physical and/or mathematical interpretations of the terms and variables in eq.~\eqref{eq:error} or \eqref{eq:error1} can be found in Table~\ref{tab:interpretation}.
\begin{table}[ht]
	\centering
	\caption{Variables and terms in eq.~\eqref{eq:error} and the corresponding specific values in eq.~\eqref{eq:error1} and the physical/mathematical interpretations.}
	\label{tab:interpretation}
	\begin{threeparttable}
		\begin{tabular}{M{1.5cm} M{1cm} M{8.2cm} M{3.8cm}}
			\toprule
			{Variables or terms} &{Value} & Mathematical/physical interpolation & Affected by \tabularnewline 
			\midrule
			$\epsilon_p$ & -    & Error field in the calculated pressure field & Everything            \\ 
			$||\epsilon_p||_{L^2(\Omega)}$ & -    &  Global measurement of the error level of the calculated pressure field & Everything            \\  
			$C_1$ & $\frac{1}{12}$      & The constant of truncation error contribution & 
			Numerical scheme \\
			$\frac{\partial^2 p }{\partial x^2}$, etc.  & - & 2nd order derivative of the pressure field  & Flow field \\
			$C_0$ &	$0.901^2$ &	Amplification ratio of the effect by Gaussian error to the ``most dangerous mode'' of the error\tnote{*} & 	Dimension, type of BCs of the domain\\
			$C_2$ &	$\frac{L^2}{2\pi^2}$ &	 Optimal Poincar\'e constant/ amplification ratio of error in the reconstructed pressure field to the error in the data of the Poisson equation & 	Dimension, area, shape, type of BCs \\
			$\sigma_u$, etc. &	- &	Variance of error of the experimental data  & 	Quality of PIV \\
			$h$ &	- &	Spatial resolution  & 	PIV experiment setup and post-processing  \\
			$m$ &	$2$ &	Scaling constant of grid spacing for the contribution from the truncation error
			& 	Numerical scheme  \\
			$n$ &	$-2$ &	Scaling constant of grid spacing for the contribution from the experimental error  & 	Numerical scheme  \\
			\bottomrule
		\end{tabular}
		\begin{tablenotes}
			\item[*] More details about the derivative, calculation, and physical interpretation of $C_0$ can be found in \cite{pan2016error2}.
		\end{tablenotes}
	\end{threeparttable}
\end{table}

The results above are developed for the non-dimensional setup. The dimensional equivalent estimates can be recovered by multiplying the variables by corresponding characteristic scales (e.g., $||\epsilon_p^*||_{L^2(\Omega)} = ||\epsilon_p||_{L^2(\Omega)} P_0$, $x^*=xL_0$, $u^*=uU_0$, etc., where $P_0$, $L_0$ and $U_0$ are characteristic pressure, length, and velocity respectively). Note that in this study the characteristic pressure is defined as $P_0=\rho U_0^2$, rather than the commonly used characteristic pressure ($P_0=\rho U_0^2/2$), where $\rho$ is the density of the fluid. In other words, the non-dimensional error level ($||\epsilon_p||_{L^2(\Omega)}$) in the current work has twice the value of the pressure coefficient ($C_p$) used in some other works (e.g., \cite{wang2017spectral}, \cite{mcclure2017optimization}). 
For convenience, the superscript ($[~]^*$) denoting dimensional variables will be dropped here after without special note and the non-dimensional variables will be written explicitly (e.g., $p/P_0$ is the non-dimensional pressure, where $p$ is the corresponding dimensional variable).

Eq.~\eqref{eq:error} can be written as a function of the spatial resolution: 
\begin{linenomath}
	\begin{equation}
	\label{eq:AnB}
	||\epsilon_p||_{L^2(\Omega)} = \text{fun}(h) \approx Ah^m + Bh^{n}, 
	\end{equation}
\end{linenomath}
where $A$ and $B$, as well as $m$ and $n$ are constants once the experimental setup, parameters, and pressure solver are determined. For example, for eq.~\eqref{eq:error1}, $A = \frac{1}{12} \Big( \left\| \frac{\partial^2 p}{\partial x^2} \right\|_{L^2(\Omega)} + \left\| \nabla^{-2}\frac{\partial^4 p}{\partial^2 x \partial^2 y} \right\|_{L^2(\Omega)} + \left\| \frac{\partial^2 p}{\partial y^2} \right\|_{L^2(\Omega)} \Big)$, $B = 0.901^2 \frac{L^2}{2\pi^2}  \left(\frac{\sigma_u^2 + \sigma_v^2}{2}\right)$, $m =2$, and $n = -2$. Clearly, eq.~\eqref{eq:AnB} is not monotonic in $h$, leaving several open questions: i) what is the minimum error ($||\epsilon_p||_{L^2(\Omega)}$)? and ii) when is the minimum achieved in terms of spatial resolution ($h$)? 

We note that $Ah^2 + Bh^{-2} \ge 2 \sqrt{AB}$, and equality is reached if and only if $Ah^2 = Bh^{-2}$, and we thus have the optimal spatial resolution 
\begin{linenomath}
	\begin{equation}
	h^{opt} \approx \sqrt[4]{B/A}, 
	\end{equation}
\end{linenomath}
which leads to an estimate of the minimum error level in the calculated pressure field: 
\begin{linenomath}
	\begin{equation}
	||\epsilon_p||_{L^2(\Omega)}^{min} \approx 2 \sqrt{AB}.
	\end{equation}
\end{linenomath}
This minimum error level can be interpreted as the overall sensitivity of the pressure reconstruction, meaning that any changes in estimated pressure smaller than this sensitivity are not physically meaningful. 
In other words, this sensitivity of the reconstructed pressure field is a global measure of the best possible accuracy of the current PIV-based pressure reconstruction. 

\subsection{Validation}
\label{sec:Derivation:Validation}
\begin{figure}[!h]
	\centering
	\includegraphics[width=0.8\textwidth]{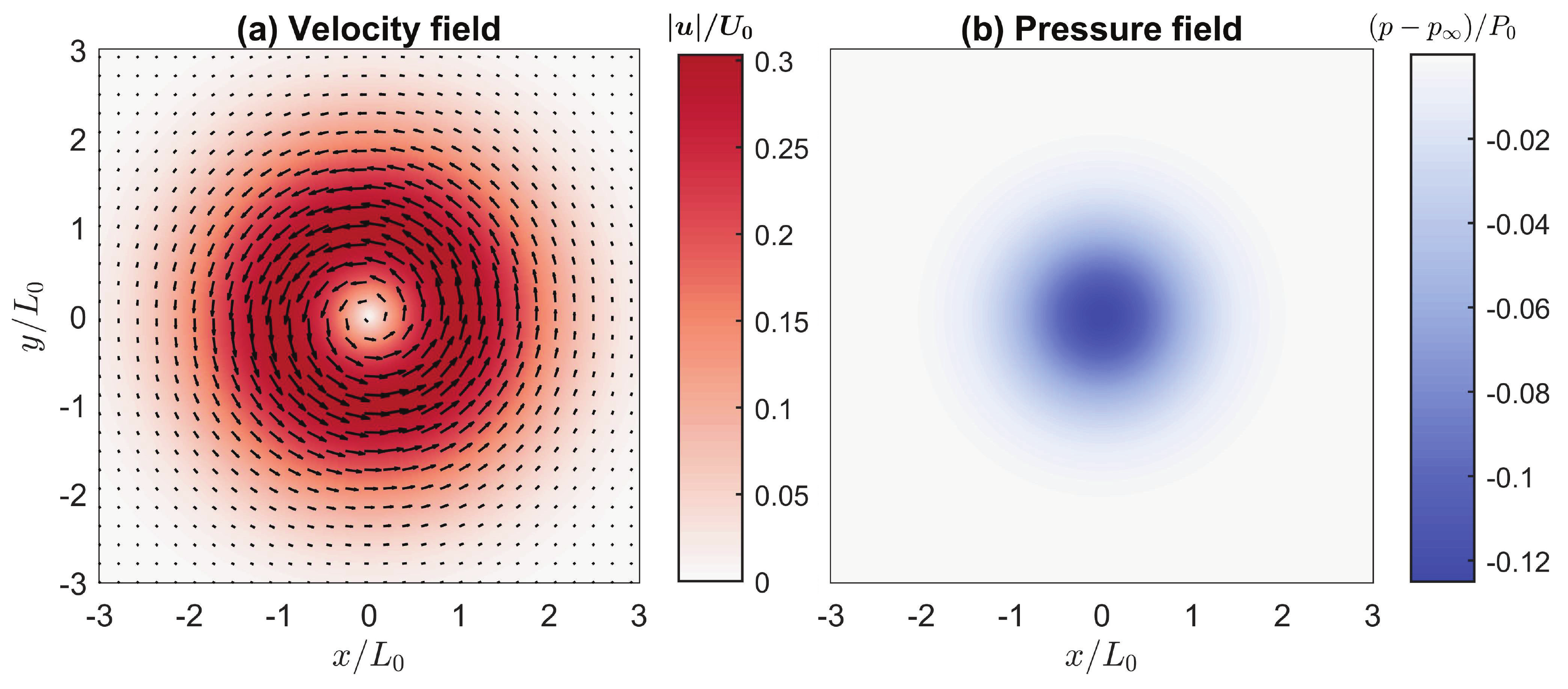}
	\caption{2D visualization of the non-dimensional flow field in a box. (a) Quiver plot of velocity field over the magnitude, and (b) the pressure field.
		\label{fig:FlowProfile}}
\end{figure}

\begin{figure}[!h]
	\centering
	\includegraphics[width=0.9\textwidth]{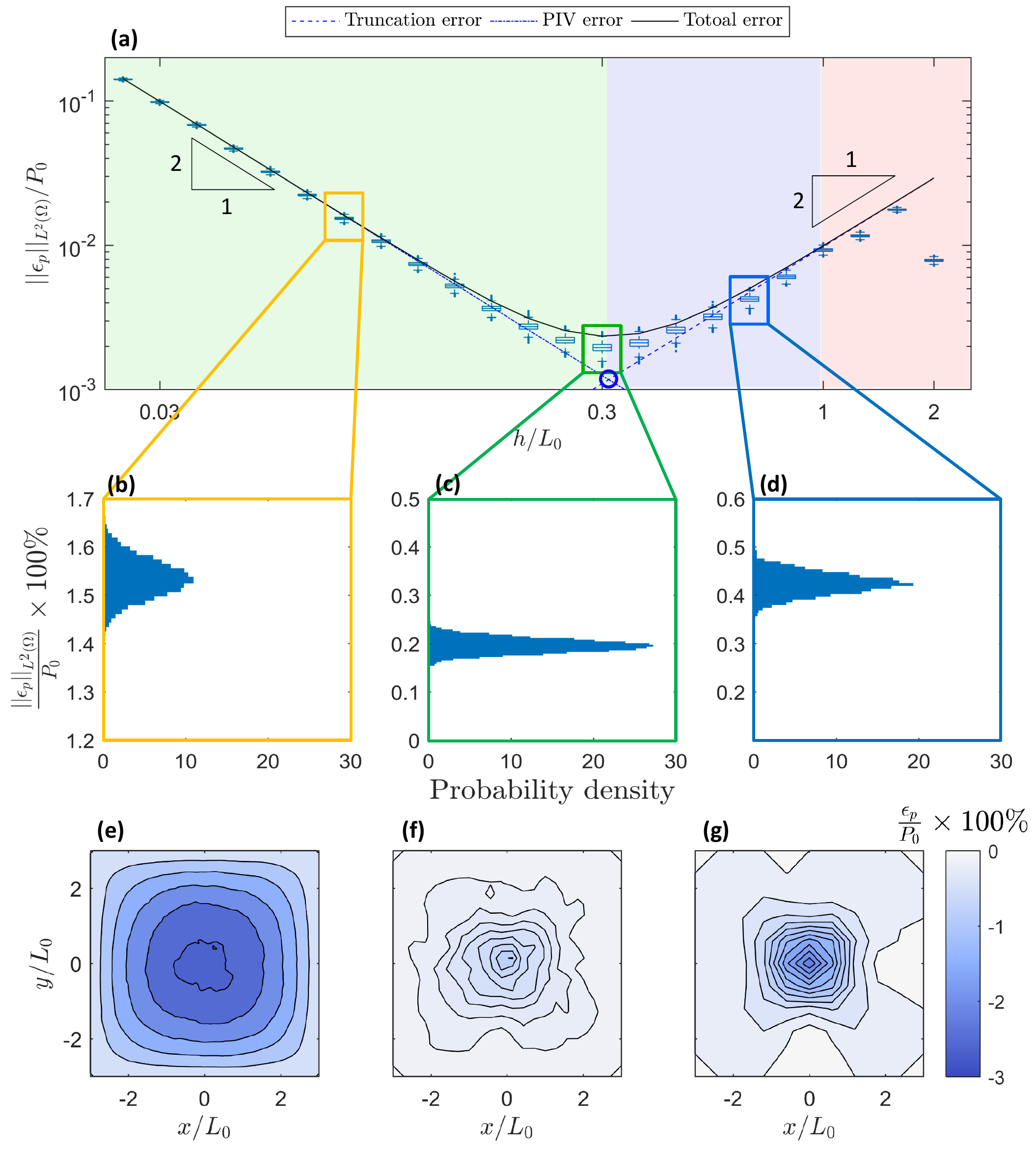}
	\caption{Error level in the calculated pressure field vs. spatial resolution. (a) Box plot of the error level in the calculated pressure field. Each box represents 5000 independent simulations. The green region is dominated by the error from the PIV measurements due to the spatial resolution being too fine. The blue region is the regime where the truncation error dominates because the resolution is too coarse. The red region indicates aliasing due to insufficient sampling lower than the Nyquist frequency. The dashed line represents the theoretical prediction of the truncation error, and the dash-dot line indicates the theoretical contribution from PIV measurement errors in the velocity field. The solid line represents the theoretical prediction of the total error in the calculated pressure field. (b-d) Normalized histograms of the relative error in the calculated pressure field for typical spatial resolutions (corresponding to the orange, green, and blue frames in Fig.~\ref{fig:BigPic}(a), respectively). (e-g) Relative error field in pressure at several spatial resolutions.}
	\label{fig:BigPic}
\end{figure}

Consider a Taylor vortex in 2D. Assuming pressure at the far field vanishes ($p_\infty=0$), the velocity and pressure fields are defined as
\begin{linenomath}
	\begin{equation}
	\label{eq:taylorU}
	u_\theta(r,t) = \frac{Hr}{8\pi \nu t^2}\exp\left(-\frac{r^2}{4\nu t}\right), 
	\end{equation}
\end{linenomath}
and 
\begin{linenomath}
	\begin{equation}
	\label{eq:taylorP}
	p_\theta(r,t) = -\rho\frac{Hr^2}{64\pi^2 \nu t^3}\exp\left(-\frac{r^2}{2\nu t}\right), 
	\end{equation}
\end{linenomath}
respectively, where $H = M / 2 \rho \nu$ is a constant that measures the amount of angular momentum $M$ in the vortex \citep{panton2006incompressible} and $M = \rho \int_{0}^{\infty} 2\pi r^2 u_\theta dr$ \citep{taylor1918dissipation}. The time is $t$, the distance from the center of the vortex is $r$, and $\rho$, and $\nu$ are density and kinematic viscosity of the fluid, respectively. We non-dimensionalize the variables as $\zeta = r/L_0$, $ \xi = u/U_0$, and $ \eta = p/P_0$, where $L_0= \sqrt{2 \nu t}$, $U_0=H/(2 \pi L_0 t)$, and $P_0= \rho U_0^2$, are the characteristic scales \footnote{These non-dimensional variables are different from the original choice from G.I. Taylor's (\citeyear{taylor1918dissipation}) similarity solutions, but more commonly used recently (e.g., \cite{trieling1998decay}), since it conserves unit vorticity at the origin and the velocity peaks when $\zeta \rightarrow 1$. More specifically, $\xi_{peak} \rightarrow \exp(-1/2)/2 \approx 0.3033$ as $\zeta \rightarrow 1$, and $p_{peak} \rightarrow -1/8=-0.125$ as $\zeta \rightarrow 0$.}. Scaling eq.~ \eqref{eq:taylorU} and \eqref{eq:taylorP} leads to
\begin{linenomath}
	\begin{equation}
	\label{eq:taylorU2}
	\xi_\theta^* = \frac{\zeta}{2}\exp\left(-\frac{\zeta^{2}}{2}\right), 
	\end{equation}
\end{linenomath}
and
\begin{linenomath}
	\begin{equation}
	\label{eq:taylorP2}
	\eta_\theta^*= - \frac{1}{8}\exp\left(-\zeta^{2}\right).
	\end{equation} 
\end{linenomath}
We will consider a ``realistic'' Taylor vortex in water with parameters shown in Table~\ref{tab:parameter}. The 2D non-dimensional representation of the flow (velocity and pressure field) is shown in Fig.~\ref{fig:FlowProfile}. We again consider point-wise Gaussian noise added to the velocity field with zero-mean and constant standard deviation (i.e., $\epsilon_u \sim \mathcal{N}(0,\sigma_u^2)$, $\epsilon_v \sim \mathcal{N}(0,\sigma_v^2)$, $\sigma_u/U_0= \sigma_v/U_0 = 7.85 \times 10^{-3})$. We refer to this numerical setup (i.e. the Taylor vortex described in Table~\ref{tab:parameter}, and this specific noise) as \textit{setup~1} hereafter. We vary the spatial resolution ($h$) of the domain and run the numerical experiments 5,000 times for each resolution. The normalized error level in the calculated pressure field ($||\epsilon_p||_{\Omega(L^2)}/P_0$) versus the normalized spatial resolution ($h/L_0$) of the domain is shown in the box plot in Fig.~\ref{fig:BigPic}(a). As mentioned, each box represents 5,000 independent numerical experiments. The theoretical predictions of the error level in the calculated pressure agree well with these numerical experiments. The blue dashed line (slope~$=2$) indicates the first term in eq.~\eqref{eq:error1}, which represents the contribution from the truncation error, which is affected by both the numerical schemes and the flow field. The blue dash-dot line (slope~$=-2$) is mainly affected by the property of the Poisson operator and the experimental noise. The black line indicates the theoretical predication of the total error (see eq.~\eqref{eq:error1}) in the calculated pressure field. The intersection of the PIV error contribution (blue dash-dot line) and the truncation error contribution (blue dashed line) is marked by the blue circle indicating the optimal spatial resolution where the minimum global error in the calculated pressure field is achieved. 

The minimum error in the calculated pressure field is $||\epsilon_p|| _{L^2(\Omega)}^{min}/P_0 \approx 2.35\times 10^{-3}$ in this specific example. For a characteristic pressure $P_0=64.85$~Pa, the best possible sensitivity of the pressure field reconstruction is approximately $0.15$~Pa. This implies that a well designed and conducted PIV experiment with an accurate pressure solver could achieve high fidelity pressure reconstructions and rival the sensitivity of pressure sensors.. Due to the `velocity-to-pressure' computation in the PIV-pressure approach, the reconstructed pressure field is scalable with the characteristic pressure ($P_0 = \rho U_0^2$). This feature indicates that PIV-based pressure reconstruction techniques are particularly attractive for applications involving small pressure changes (e.g., slow air flows introduce relatively low values of $\rho$ and $U_0$, and thus low $P_0$), which often requires high cost instrumentally when using high-sensitivity pressure transducer arrays. For example, assuming an air flow having the same velocity field as \textit{setup~1}, the low density of the fluid media (e.g., $\rho \approx 1$~$\text{kg/m}^3$) leads to a low characteristic pressure ($P_0 \approx 0.065$~Pa). 
The corresponding pressure measurement sensitivity in such a PIV-pressure calculation can be approximately as high as $\sim 1.5 \times 10^{-4}$~Pa. 
Therefore, in addition to the ability to measure pressure anywhere in a flow field, pressure from PIV has the potential for superior accuracy for slow flows.

Remembering the dynamic range ($D$) is the ratio between the maximum measurable ($p^{max}$) and the sensitivity ($||\epsilon_p|| _{L^2(\Omega)}^{min}$), we define the dynamic range of the PIV-based pressure calculation techniqes in the current paper as 
\begin{linenomath}
	\begin{equation}
	\label{eq:DR}
	D = \frac{p^{max}}{||\epsilon_p|| _{L^2(\Omega)}^{min}}.
	\end{equation} 
\end{linenomath}
We expect that PIV-based pressure calculation techniques have the following features: i) the maximum measurable pressure is determined by the velocity field and the fluid density, and is scalable to $\rho U_0^2$. In other words, $p^{max}$ is flow dependent; ii) noting that the sensitivity of the measurement is affected by many factors (see Fig.~\ref{fig:puzzle} and eq.~\eqref{eq:error}), the sensitivity is not a fixed value either; 
iii) thus, the PIV-based pressure reconstruction techniques have a `\textit{dynamic}' dynamic range, which depends on many factors including the nature of the flow. This property is distinct to conventional pressure transducers' \textit{fixed} dynamic range. The dynamic range of the PIV-based pressure reconstruction could be high if the experiment and pressure solver are carefully designed. For example, in the example presented above, the dynamic range is $D=4.3 \times 10^{5}$, which is comparable to (or even greater than) current typical pressure gauges.


Since the contribution from truncation error scales as $||\epsilon_{p,T}||_{L^2(\Omega)} \sim O(h^{2})$, and the contribution from measured noise in the velocity field scales as $||\epsilon_{p,E}||_{L^2(\Omega)} \sim O(h^{-2})$, we expect there to be a competition between the two errors in terms of the resolution $h$.
This phenomena has been observed previously \cite{charonko2010assessment,mcclure2017optimization,pan2016errorT}, as well as in the current study (e.g., Fig.~\ref{fig:BigPic}(a)). 
In the following we provide an explicit and accurate interpretation based on a rigorous analysis (e.g., eq.~\eqref{eq:error}). 
When the spatial resolution is too small (e.g., $h/L_0 \rightarrow 0$), the error in the pressure is dominated by the error from the noise in the velocity field (green patched regime in Fig.~\ref{fig:BigPic}(a)). 
When the spatial resolution is relatively large (e.g., $h/L_0 \rightarrow 1$), the spatial resolution is comparative to the length scale of the flow structure, and the truncation error due to the discrete scheme is the dominant error source (blue patched regime). 
When the spatial resolution is even larger (e.g., $h/L_0 \gg 1$), the insufficient sampling lower than the Nyquist frequency causes aliasing and unreliable, or more precisely, meaningless pressure field reconstructions. 

\begin{table}[ht]
	\centering
	\caption{Parameter space for a numerical experiment}
	\label{tab:parameter}
	\begin{tabular}{M{6cm} M{5cm} M{1.5cm}}
		\toprule
		\textbf{Parameters} & \textbf{Value} & \textbf{Unites} \tabularnewline 
		\cmidrule{2-3}
		$H$ & $10^{2}$ & $[\text{m}^2]$            \\
		$\rho$ & $10^{3}$ & $[\text{kg/m}^{3}]$             \\
		$\nu$ & $10^{-6}$ & $[\text{m}^2/\text{s}]$             \\
		$t$ & $1,250$ & $[\text{sec}]$             \\ 
		\midrule
		\textbf{Characteristic scales} & ~   & ~ \\
		$L_0= \sqrt{2\nu t}$ & 0.05   & [m] \\
		$U_0=\frac{H}{2\pi L_0 t}$ & 0.255   & [m/s] \\
		$p_0= \rho U_0^2$ & 64.85    & [Pa] \\
		\midrule
		\textbf{Dimensionalized variables} & ~   & ~ \\
		$[x \times y]$ & $[-0.15, 0.15] \times [-0.15,0.15]$   & [m] \\
		$u_{peak}$ & 0.0772    & [m/s] \\
		$p_{peak} - p_\infty$ & $-8.106$    & [Pa] \\
		\midrule
		\textbf{Non-dimensionalized variables} & ~   & ~ \\
		$[x \times y]/L_0$ & $[-3, 3] \times [-3,3]$   & - \\
		$u_{peak}/U_0$ & 0.3033    & - \\
		$(p_{peak} - p_\infty)/P_0$ & $-0.125$    & - \\
		\bottomrule
	\end{tabular}
\end{table}

The normalized histograms of the error level in the calculated pressure field at three different values of the spatial resolution as indicated in Fig.~\ref{fig:BigPic}(a) (marked by orange, green, and blue frames), are shown in Fig.~\ref{fig:BigPic}(b-d), respectively. These histograms are normalized by $||\epsilon_p||_{L^2 (\Omega) }/P_0 \times 100\%$. They represent the probability density function (PDF) of the relative error (percentage compared to the characteristic pressure). One of the error fields in the reconstructed pressure drawn from the 5,000 independent numerical experiments for the three typical spatial resolutions are shown in Fig.~\ref{fig:BigPic}(e-g), respectively.


We note that point-wise Gaussian noise in the velocity field led to an error level in the pressure field with a near Gaussian distribution (the histograms in Fig.~\ref{fig:BigPic}(b-d) appear Gaussian). This ``Gaussian-input Gaussian-output'' property would be expected for a linear transformation, but the pressure construction is a highly nonlinear process in general~\footnote{Although the influence of the data on the pressure ($f \rightarrow p$) through the Poisson equation is a linear process, the nonlinear transformation from the velocity to the data ($\bm{u} \rightarrow f$) makes the error propagation process nonlinear.}. 
A heuristic explanation could be that the Poisson equation based pressure solver is well-approximated by a linear transformation over the small range of the noise, and that a noise of larger variance would be needed to observe nonlinear effects. 
A precise description of this approximation is an open question that we will consider in future work.

More general validations can be achieved by varying the error level in the velocity field (e.g., different $\sigma_u^2$ and $\sigma_v^2$) and adjusting the flow field (e.g., a flow with different characteristic scales). We consider i) the same flow used in the above example (see Table~\ref{tab:parameter} and Fig.~\ref{fig:FlowProfile}), but with larger error with different statistics (i.e., $\epsilon_u \sim \mathcal{N}(0,\sigma_u^2 ), \epsilon_v \sim \mathcal{N}(0,\sigma_v^2 )$, where $\sigma_u/U_0 = 1.57 \times 10^{-2}$ and $\sigma_v/U_0 = 3.93 \times 10^{-3}$, called \textit{setup~2} hereafter); ii) the younger stage ($t=312.5$~sec) of the same decaying vortex (see Table~\ref{tab:parameter2} for detailed parameters) in the same dimensional domain (meaning a larger non-dimensional size of the domain), and the same dimensional error level as \textit{setup~1} (i.e., $\epsilon_u \sim N(0,\sigma_u^2), \epsilon_v \sim N(0,\sigma_v^2)$, where $\sigma_u /U_0=\sigma_v /U_0 = 0.98 \times 10^{-3}$, called \textit{setup~3} hereafter). 

\begin{table}[ht]
	\centering
	\caption{Parameters of two different flows for validation (setup~1 \& 2, and a younger vortex for setup~3). }
	\label{tab:parameter2}
	\begin{tabular}{M{3cm} M{3cm} M{2cm} M{1cm}}
		\toprule
		\textbf{Parameters} & \textbf{Setup~1~\&~2} & \textbf{Setup~3}  & \textbf{Units} \tabularnewline 
		\midrule
		$L_0= \sqrt{2\nu t}$ & 0.05    & 0.025 &  [m]\\
		$U_0=\frac{H}{2\pi L_0 t}$ & 0.25    & 2.04  & [m/s]\\
		$P_0= \rho U_0^2$ & 64.85    & 4,150  & [Pa]\\
		$u_{peak}$ & 0.1   & 0.87  & [m/s]\\
		$p_{peak} - p_\infty$ & $-8.1$  & $-518.8$  & [Pa]\\
		Re &	$~1.27 \times 10^4$ &	$~5.0\times 10^4$ & - \\
		\bottomrule
	\end{tabular}
\end{table}

\begin{figure}[!h]
	\centering
	\includegraphics[width=0.8\textwidth]{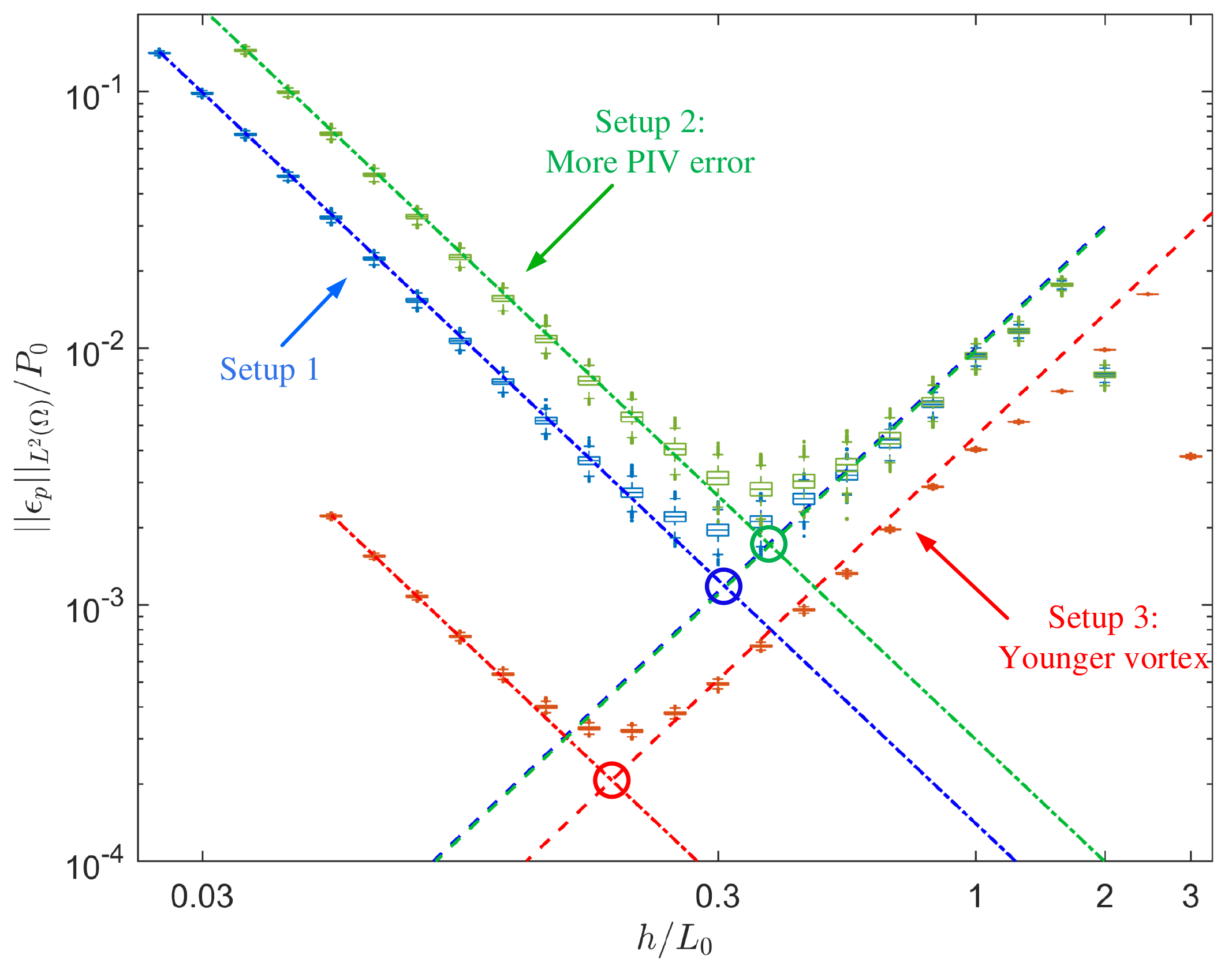}
	\caption{Nondimensional error in the calculated pressure field vs. non-dimensional spatial resolution. Numerical experiments of \textit{setup~1} (blue), \textit{setup~2} (green), and \textit{setup~3} (red). The dashed lines indicate the contribution from truncation error, and the dash-dot lines indicate the contribution of the error from the PIV measurement in the velocity field. The optimal spatial resolutions are marked by the circles at the intersections of the dashed lines and the dash-dot lines, with corresponding color schemes.}
	\label{fig:3setups}
\end{figure}

\begin{figure}[!h]
	\centering
	\includegraphics[width=0.8\textwidth]{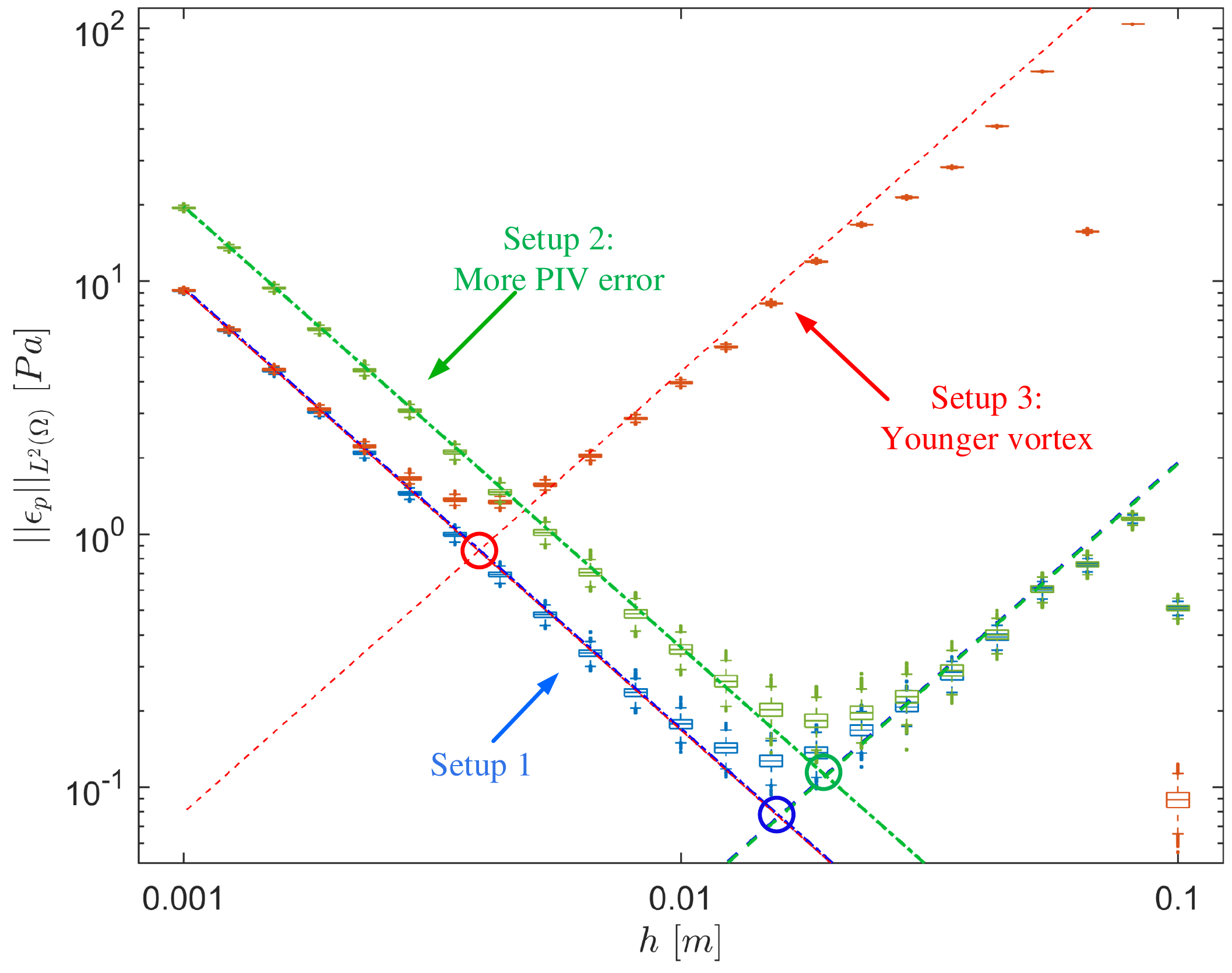}
	\caption{Error in the calculated pressure field vs. spatial resolution. Numerical experiments of \textit{setup~1} (blue), \textit{setup~2} (green) and \textit{setup~3} (red). The dashed lines indicate the contributions from the truncation errors, and the dash-dot lines indicate the contribution of the error from the PIV velocity measurement. The optimal spatial resolution is marked by the circles on the intersections of the dashed lines and dash-dot lines, with corresponding color schemes.}
	\label{fig:3setups2}
\end{figure}

Similar numerical experiments are conducted and the results are shown in Fig.~\ref{fig:3setups}. 
The results from the numerical experiments agree with the theoretical predictions well for various flows and PIV error statistics. 
Comparing \textit{setup~1} and \textit{setup~2}, which share the same flow field but different error statistics in the velocity field, we note that when the spatial resolution is large (e.g., larger than the optimal resolution of the \textit{setup~2}, marked by the green circle in Fig ~\ref{fig:3setups}), the truncation error dominates and the numerical experimental results collapse onto the same dashed line, which is solely determined by the nature of the flow. 
When the spatial resolution is small, the error in the velocity field from the PIV measurements is the major contributor to the error in the calculated pressure field. 
Thus, \textit{setup~2} introduces more error than \textit{setup~1}, and the optimal spatial resolution is coarser than it is for \textit{setup~1} (the green circle is on the right of the blue circle). 
Comparing \textit{setup~1} and \textit{setup~3}, a smaller characteristic length (radius of the vortex) of the flow in \textit{setup~3} implies that the optimal resolution for \textit{setup~3} is finer than \textit{setup~1} or \textit{setup~2} since a smaller scale flow structure must be resolved (the red circle is on the left of the green and blue circles). 
We emphasize that the vertical axis in Fig.~\ref{fig:3setups} is a non-dimensional error level, not the dimensional value. 
Instead, $||\epsilon_{p}||_{L^2(\Omega)}/P_0$ is an `error level' comparing the error to the corresponding characteristic pressure. 
Noting that \textit{setup~3} has significantly higher characteristic pressure than \textit{setup~1} and \textit{setup~2}, it is not surprising that the minimum error level (or sensitivity of the pressure measurement) for \textit{setup~3} is lower than the other two setups (the red circle is located lower than the green and blue ones). 
However, this does not necessarily mean that the absolute pressure sensitivity for \textit{setup~3} is low. 
A more intuitive presentation can be found in Fig.~\ref{fig:3setups2}, which is reconstructed from Fig.~\ref{fig:3setups}, but with physical dimensions included.

Figure~\ref{fig:3setups2} shows the error in the calculated pressure field versus spatial resolution for \textit{setup~1} (blue), \textit{setup~2} (green), and \textit{setup~3}(red). 
When the spatial resolution is small (e.g., to the left of the red circle in Fig.~\ref{fig:3setups2}), the numerical experimental results (blue boxes and the red boxes) are collapsed onto the same dash-dot line because the same error statistics are shared as well as the same domain properties (e.g., size of the domain, type of BCs, etc.). 
The error in \textit{setup~2} is higher than that from \textit{setup~1} and \textit{setup~3} when the spatial resolution is small due to the larger random noise in the velocity field. 
When the spatial resolution is high (e.g., to the right of the green circle), the numerical experimental results from \textit{setup~1} and \textit{setup~2} (blue and green boxes) are collapsed onto the same theoretical prediction since they have the same flow field, despite these two setups having different noise statistics in the velocity field.

More importantly, Fig.~\ref{fig:3setups2} clarifies how the flow field and error in the PIV measurements affect the optimal spatial resolution and the pressure reconstruction sensitivity (note the vertical positions of the colored circles) for the three different setups. 
The larger error in the PIV measurements requires coarser optimal spatial resolution and leads to lower pressure reconstruction sensitivity (comparing the positions of the blue and green circle). 
The smaller dominant flow structures in the flow require finer spatial resolution, however, leading to worse minimum resolvable pressure (comparing the positions of blue and the red circle).

\begin{figure}[!h]
	\centering
	\includegraphics[width=0.9\textwidth]{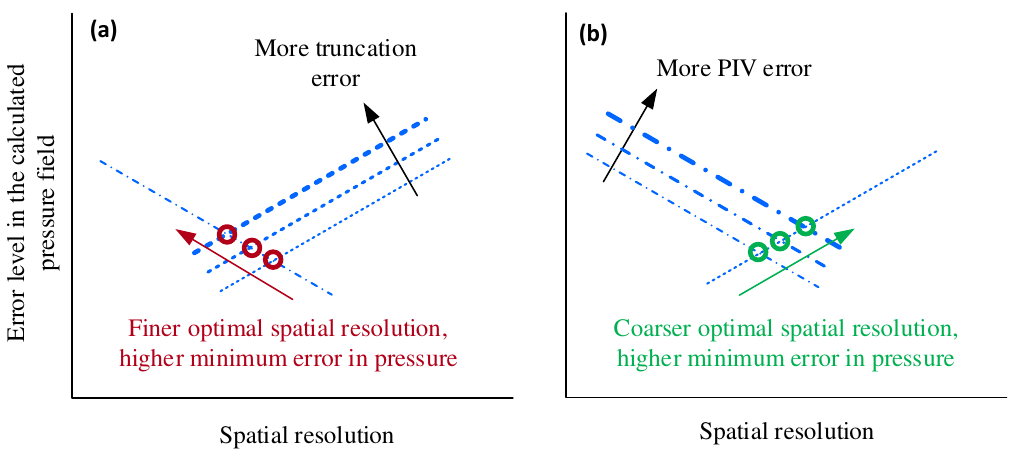}
	\caption{Qualitative illustration of the contributions and/or competition of the truncation error and PIV error. More truncation error in the domain leads to finer optimal spatial resolution, and higher minimum error in the calculated pressure field (marked by the red circles and arrow head in (a)). More error from the PIV experiments leads to coarser optimal resolution and higher minimum error in the pressure field.}
	\label{fig:Cartoon}
\end{figure}

A qualitative illustration of how the error from the PIV experimental measurement and the truncation error from the numerical solver compete against each other for the optimal spatial resolution, and at the same time, contribute together to the minimum error in the pressure field is shown in Fig.~\ref{fig:Cartoon}. 
Larger truncation error (e.g. due to a flow with higher spatial frequency) would shift the dashed lines up (Fig.~\ref{fig:Cartoon}(a)) and lead to a requirement for finer spatial resolution to achieve the minimum error in the pressure field (see the locus marked by the red circles and arrow head in Fig.~\ref{fig:Cartoon}(a)).
More error in the velocity field from the PIV experiments will shift the dash-dot line up and require coarser spatial resolution for the minimum error in the calculated pressure field (see the locus marked by the red circles and arrow head in Fig.~\ref{fig:Cartoon}(b)). 
Based on the above observations, an intuitive impression is that one of the most challenging PIV experimental results for PIV-based pressure reconstruction is a flow with small scale dominant structures (usually leading to small characteristic length scales and more significant contributions from the truncation error) and high uncertainties in the velocity field. 

\section{A practical method for estimating error}
\label{sec:Practical}

Although the current research mainly focuses on theoretical insights of the resolution of the reconstructed pressure and the corresponding optimal spatial resolutions, we will also  provide brief guidelines for engineering practices. 
In addition, we propose a practical methodology to estimate the errors in the pressure field.

The theoretical prediction of an optimal spatial resolution and sensitivity of the pressure reconstruction involves careful estimates of some constants (see eq.~\eqref{eq:error}), such as the optimal Poincar\'e constants ($C_2$) and the amplification ratio $C_0$, for the PIV-noise dominant regime, as well as 
the constants derived from the corresponding Taylor series to estimate the truncation error. 
These constants are not trivial to compute. 
Of these constants, the Poincar\'e constant is the most amenable to finding the exact value, but even then, for an arbitrary domain the Poincar\'e constant is likely best estimated via the Rayleigh quotient, which is an expensive calculation.
Here we provide a practical method for estimating these constants. 
Eq.~\eqref{eq:error} can be written as 
\begin{linenomath}
	\begin{equation}
	\label{eq:error2}
	\| \epsilon_{p}\|_{L^2(\Omega)} \approx K_1 h^2 + K_2 (\frac{\sigma_u^2+ \sigma_v^2}{2}) h^{-2},
	\end{equation}
\end{linenomath}
where $K_1h^2$ is the truncation error contribution and supposedly requires \textit{a priori} information of the true value of the pressure field (see eq.~\eqref{eq:error}). The constant
$K_2=C_0C_2$ reflects a compound amplification ratio due to the setup of the pressure solver and the profile of the error in the PIV data. The purpose of the following work is to estimate $K_1$, $K_2$, and in turn $||\epsilon||_{L^2(\Omega)}$, using the measured velocity $\tilde{\bm{u}}$ and the corresponding uncertainty quantification (e.g., $\sigma_u$ and  $\sigma_v$) alone without any knowledge of the true values of the velocity field $\bm{u}$ and pressure field $p$.

When the grid spacing $h$ is sufficiently large, the truncation error dominates and eq.~\eqref{eq:error2} reduces to 
\begin{linenomath}
	\begin{equation}
	\label{eq: error est K1}
\| \epsilon_{p}\|_{L^2(\Omega)}   =  \| \epsilon_{p}\|_{L^2(\Omega)}(h)\approx K_1 h^2,
	\end{equation}
\end{linenomath}
and $K_1$ can be estimated using any standard grid convergence criteria from numerical analysis. 
Here, we adopt Richardson extrapolation \citep{zhang2021numerical, roache1993completed} for this purpose. 
Assuming the PIV results are obtained at a certain resolution with grid spacing $h$, a direct pressure field reconstruction based on this data is $\tilde p(h)$.
Deceasing the spatial resolution by down-sampling leads to a coarser grid spacing $r_1h$ ($r_1>1$), and the calculated pressure on the down-sampled data is denoted as $\tilde p(r_1h)$, where $r_1$ is the down-sampling factor,  
The error in $\tilde p(h)$ can be estimated as
\begin{linenomath}
	\begin{equation}
	\label{eq:Richardson}
	\hat{\epsilon}_p(h) = -\frac{\tilde{p}(h)-\tilde{p}(r_1h)}{r_1^q-1},
	\end{equation}
\end{linenomath}
where $q$ is the order of accuracy of the discretization scheme. 
In the context of the current research,
$q=2$ due to the use of a second-order central differencing method (i.e.,  eq.~\eqref{Aeq:PartialError}).
By setting $r_1$ to 2, $\epsilon_p(h)$ and its norm $||\epsilon_p||_{L^2(\Omega)}(h)$ can be calculated with various gird spacing $h$. 
Invoking eq.~\eqref{eq: error est K1}, $K_1$ can be found by linear regression of $||\epsilon_p||_{L^2(\Omega)}(h)$ against $h^2$.

It is important to note that, the above estimation only applies to the truncation error dominated regime and would fail to estimate the contribution from PIV error. 
In the PIV error dominated regime, the contribution from truncation error is negligible and eq.~\eqref{eq:error2} can be reduced to 
\begin{linenomath}
	\begin{equation}
	\label{eq: error est K2}
	\| \epsilon_{p}\|_{L^2(\Omega)} = 	\| \epsilon_{p}\|_{L^2(\Omega)}(\sigma_{u,v}) \approx  K_2 (\frac{\sigma_u^2+ \sigma_v^2}{2}) h^{-2}.
	\end{equation}
\end{linenomath}
We estimate $K_2$ by introducing artificial `overwhelming noise' to the measured velocity field. 
The overwhelming noise has two features: 
i) The amplitude of the noise must be significantly higher than the inherent errors of the experimental data so that the added artificial noise saturates the unknown but smaller noise in the original PIV data; 
ii) the noise is artificially assigned (e.g., white noise with variance $S_{u,v} \gg \sigma_{u,v}$) thus the statistics of the noise are explicitly known.
Adding the artificial overwhelming noise in the original velocity field will significantly corrupt the calculated pressure field (denoted as ${p}(S_{u,v})$), meaning that ${\epsilon}_{p}(S_{u,v})$ would be much larger than the actual but unknown error in the pressure field reconstructed based on the original PIV data (i.e. ${\epsilon}_{p}(S_{u,v}) \gg {\epsilon}_{p}(\sigma_{u,v}) $).
In turn, the corresponding norm of the error is also larger: 
$\| \hat{\epsilon}_{p}\|_{L^2(\Omega)} (S_{u,v}) \gg \| \hat{\epsilon}_{p}\|_{L^2(\Omega)} (\sigma_{u,v})$. 
Thus, the error level in the calculated pressure field when artificial overwhelming noise $S_{u,v}$ is used, becomes 
\begin{linenomath}
	\begin{equation}
	\label{eq:additional error}
	\| \hat{\epsilon}_{p}\|_{L^2(\Omega)} (S_{u,v}) \approx \| \tilde{p}(S_{u,v})-\tilde{p}(\sigma_{u,v})\| _{L^2(\Omega)},
	\end{equation}
\end{linenomath}
where $\tilde{p}(\sigma_{u,v})$ refers to the calculated pressure field without artificial overwhelming noise added.
Keep in mind that our goal is to estimate $K_2$, which is invariant once the fundamental setup of the experiment is established.
Note that eq.~\eqref{eq:additional error} can be viewed as a function of $S_{u,v}$. 
We can then estimate $K_2$ by varying the level of the artificial overwhelming noise, which is scaled up by $r_2$ times that of the intrinsic error in the original PIV data ($S_{u,v} = r_2 \sigma_{u,v}$ and $r_2 \gg 1$ is a scaling factor).
When $r_2$ is large, eq.~\eqref{eq:additional error} becomes
\begin{linenomath}
	\begin{equation}
	\label{eq:overwhelming}
	\| \hat{\epsilon}_{p}\|_{L^2(\Omega)} (S_{u,v}) \approx  K_2  (\frac{S_u^2+ S_v^2}{2}) h^{-2}  =  K_2 r_2^2 (\frac{\sigma_u^2+ \sigma_v^2}{2}) h^{-2}.
	\end{equation}
\end{linenomath}
The intrinsic statistics of the error in the PIV data, $\sigma_u$ and $\sigma_v$, can be estimated by uncertainty quantification techniques such as those by \cite{wieneke2015piv,sciacchitano2016piv,sciacchitano2019uncertainty}. 
By varying $r_2$, linear least squares regression can be applied to eq.~\eqref{eq:overwhelming} to estimate the coefficient $K_2$.

\begin{figure}[!h]
	\centering
	\includegraphics[width=1\textwidth]{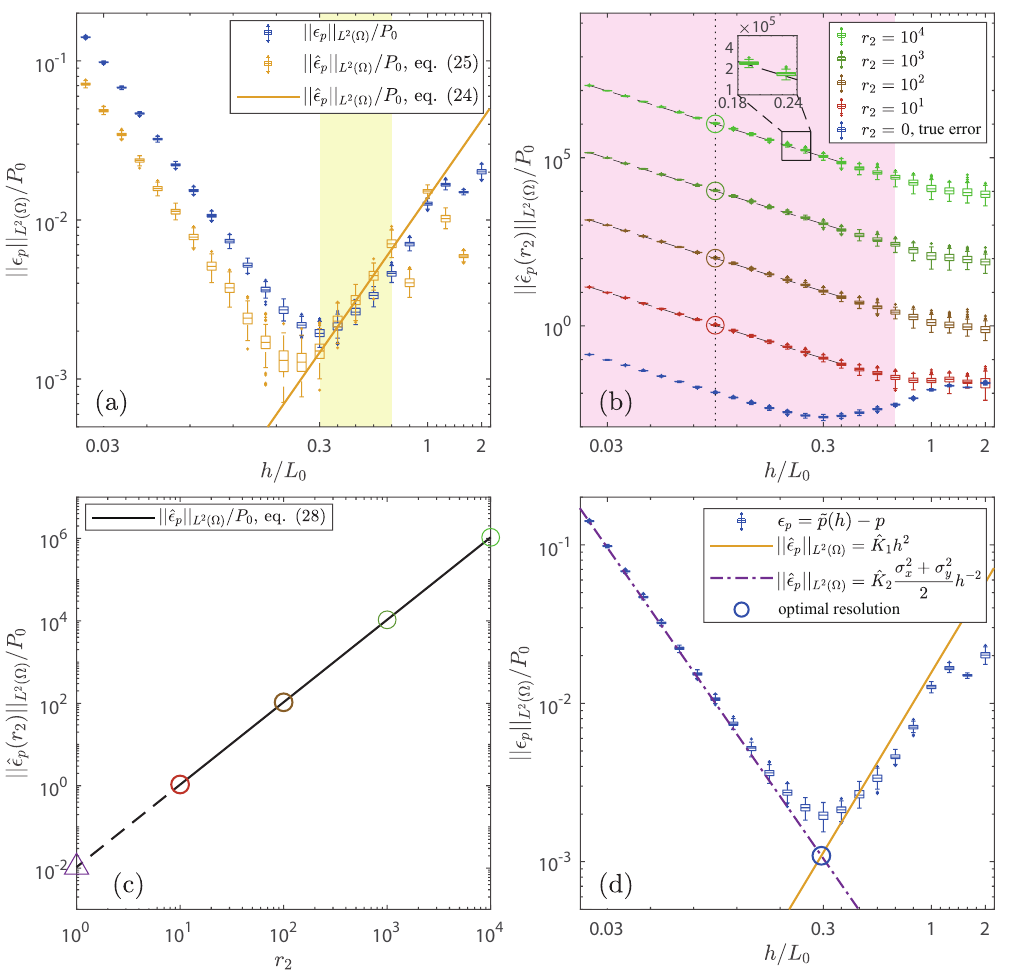}
	\caption{(a) Error level in the calculated pressure field  $||\epsilon||_{L^2(\Omega)}$ estimated by Richardson extrapolation. Blue and orange boxes are the true error and  estimated error at various grid spacing $h/L_0$, respectively. The orange line represents the fitting of eq.~\eqref{eq: error est K1} using the yellow boxes located in the light yellow patch.
	(b)~Error in the calculated pressure field when various artificial overwhelming noise is added to the velocity field (all colored boxes except blue). Blue boxes are the true error without additional noise added, which corresponds to \textit{setup~1} in Fig.~\ref{fig:3setups}. 
	Dashed lines are the estimated error based on eq.~\eqref{eq:overwhelming} fitted by linear regression. Dotted lines refer to $h/L_0=0.09$, where the added error is marked by colored open circles.
	(c)~Linear regression of error in the pressure field (marked by the circles with the same color as in (c)) that allows calculation of $K_2$ in eq.~\eqref{eq:overwhelming} and the extrapolation to the true error (purple triangle) at $h/L_0=0.09$. 
	(d) The comparison of estimated error by eq.~\eqref{eq: error est K1}, eq.~\eqref{eq: error est K2} and the true error. The blue circle marks the optimal resolution. 
		\label{fig:Kestimated}}
\end{figure}

Now we use \textit{setup~1} to validate the proposed estimation methods above for estimating $K_1$ and $K_2$.
In Fig.~\ref{fig:Kestimated}(a), orange boxes represent the estimated error based on Richardson extrapolation by setting $r_1=2$ in eq.~\eqref{eq:Richardson}.
As  eq.~\eqref{eq:Richardson} aims to predict the truncation error only and is thought to scale with $||\hat{\epsilon}_{p}||_{L^2(\Omega)}\sim h^2$, we only use the boxes with a trend of slope of two (see the boxes in the yellow patch in Fig.~\ref{fig:Kestimated}) to fit a straight line (the orange line in Fig.~\ref{fig:Kestimated}) to estimate $K_1$. 
In this test case, $K_1 = 1.82\times10^{-2}$ with coefficient of determination $R^2 = 0.97$.
For comparison, we also plot blue boxes in Fig.~\ref{fig:Kestimated} to show the true value of the error in the pressure field, which is the same as the blue boxes as shown in Fig.~\ref{fig:3setups2}. 
It is interesting to note that the error estimate based on eq.~\eqref{eq:Richardson} generally follows the trend of the true error even when the grid spacing is small despite the significant underestimation. 

To estimate $K_2$, we vary $r_2$ so that $r_2 = 10, 10^2, 10^3,~\text{and}~10^4$ in eq.~\eqref{eq:overwhelming}. 
Boxes of the same color represent the estimate of the error in the pressure field caused by the same level of  added overwhelming noise in the PIV results.
As shown in Fig.~\ref{fig:Kestimated}(b), when the grid spacing $h$ is relatively small, each group of the estimated errors followa a line of slope of $-2$, no matter how large the added artificial noise is (recall eq.~\eqref{eq:overwhelming}). 
Linear regression for each group of estimated errors in the pressure field on logarithm scale finds the dashed line.
These lines are at different heights due to different levels of added noise. 
The fitted  (dashed black line) and the estimated error (green boxes, as an example) are illustrated in the inset box in Fig.~\ref{fig:Kestimated}(b). 
For a given sufficiently small resolution\footnote{Grid spacing $h/L_0$ locates in the regime where the boxes with the same color having a slope of $-2$ (e.g., the red patch in Fig.~\ref{fig:Kestimated}(b))}, for example, $h/L_0=0.09L_0$ as shown by the vertical dashed line in Fig.~\ref{fig:Kestimated}(b), we can find the estimated error in the pressure field at this resolution (marked by the colored circles in Fig.~\ref{fig:Kestimated}(b)) for different added overwhelming noise.
Plotting these errors $\|\hat{\epsilon}\|_{L^2(\Omega)}$ against the scaling factor $r_2$ at the constant resolution $h$, we expect that the data (i.e., colored circles in Fig.~\ref{fig:Kestimated}(c), which also correspond to the ones in Fig.~\ref{fig:Kestimated}(b)) to align with a straight line with a slope of 2 according to eq.~\eqref{eq:overwhelming} subject to the constant $K_2$. 
The black solid line in Fig.~\ref{fig:Kestimated}(c) is the linear regression for $K_2$. 
In this test case, we estimate $K_2 = 1.47$, which is close to the exact value ($K_2 = 1.48$) as show in Table~\ref{tab:interpretation}. 
The extrapolation of the fitted line is expected to pass the true error level in the pressure field, as marked by the purple triangle in  Table~\ref{tab:interpretation}. 

With $K_1$ and $K_2$  estimated, we can reproduce the whole error estimate for the pressure field.
As shown in Fig.~\ref{fig:Kestimated}(d), the purple dash-dotted line on the left and the orange solid line on the right are the error estimates based on eq.~\eqref{eq: error est K1} and eq.~\eqref{eq: error est K2}, respectively. 
The intersection of these two lines indicates the optimal resolution and an estimate of the minimal error of pressure calculation, which agree well with the true error in the pressure field.


\section{Effect of interrogation window overlap and correlated error}
\label{sec:Practical:Overlap}
The above analysis and results are based on the assumption that the error or noise at neighboring PIV vectors are independent. 
This point-wise independence assumption is not required.
Overlapped interrogation windows are a common practice for PIV processing to enhance the measurement quality and increase spatial resolution.  Interrogation window overlap will introduce correlation to both the `signal' and the error at neighbouring locations.
This correlation of the error may in turn influence the error propagation.
In this section, we briefly discuss how the correlated error (due to an overlapped interrogation window) at nearby vectors may influence the error propagation from the velocity field to the reconstructed pressure field.

The dominant pathway that error propagates from the velocity field to the pressure field is through the positive definite terms~(see eq.~\eqref{Aeq:E}).
The expected value of these terms, for example $\mathbb{E}\left[\left(\left.{\partial{\epsilon_u}}/{\partial x}\right|_{i,j}\right)^2\right]$ can be affected by the overlap ratio of the interrogation window (detailed derivation can be found in Appendix~\ref{sec:Appendix overlap}):
\begin{linenomath}
	\begin{equation}
	\begin{aligned}
    \label{eq:MeanVarOv}
	\mathbb{E}\left[\left(\left.\frac{\partial{\epsilon_u}}{\partial x}\right|_{i,j}\right)^2\right] =
    \begin{cases}
        (1 - ov)\dfrac{h^2}{\sigma_u^2} , & ov \leq 50\% \\
    \dfrac{\sigma_u^2}{2h^2}, & 50\%<ov\leq100\%
    \end{cases}
	\end{aligned}
	\end{equation}
\end{linenomath}
Thus the error estimate in eq.~\eqref{eq:error} will also necessarily be altered.



\begin{figure}[!h]
	\centering
	\includegraphics[width=0.9\textwidth]{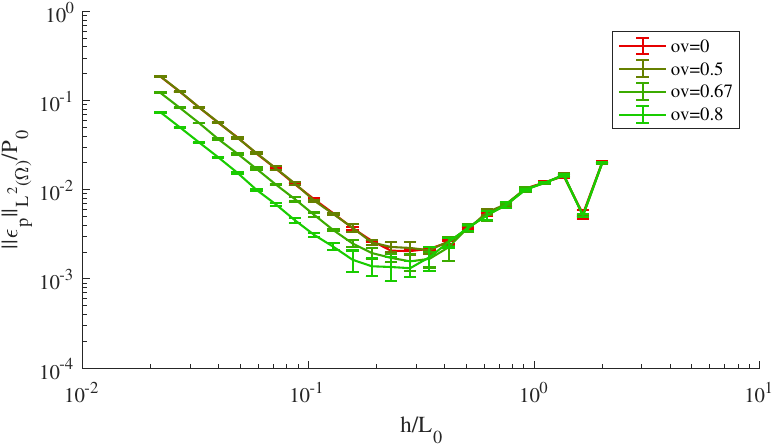}
	\caption{Error level of the calculated pressure field $||\epsilon||_{L^2(\Omega)}$ for various grid spacing $h/L_0$ for simulated PIV data with different overlap ratio. The blue boxes are corresponding to \textit{setup~1} in Fig.~\ref{fig:3setups} and \ref{fig:Kestimated} Dashed lines are the curves fitted based on the data boxes. }
	\label{fig:Overlap}
\end{figure}

For validation, \textit{setup~1} is used  again to test how the interrogation window overlap ratio and the corresponding correlated error may affect the error propagation.
In this test, PIV data generated using four common overlap ratios (i.e., 0\%, 50\%, 75\%, 87.5\%) are employed to calculate the pressure field.
The error level in the pressure field against spatial resolution are plotted as boxes in Fig.~\ref{fig:Overlap} for different interrogation window overlap ratios. 
The general trend of the boxes is similar as seen in previous validation tests.  
This implies that the influence of the interrogation window overlap in pressure calculations is rather limited.
When the grid spacing $h/L_0$ is large, as shown in the right zoomed-in window in Fig.~\ref{fig:Overlap}, the boxes for different $ov$ collapse, which is consistent with the conclusion that the integration window overlap only affects the PIV error term instead of the truncation error term (see eq.~\eqref{eq:error}). 
As shown in the left zoomed-in window in Fig.~\ref{fig:Overlap}, the boxes for $ov=0\%$ coincides with that for $ov=50\%$, which justifies the conclusion in eq.~\eqref{eq:MeanVarOv} suggesting that an overlap ratio less than 50\% would not affect the error propagation.\footnote{Note, this is a result based on the use of the central difference scheme and other numerical implementation of the pressure solver many change this result.} 

When $h/L_0$ is small, higher $ov$ reduces the error in the calculated pressure field. 
For example, the fitting lines for each group of the boxes corresponding to different $ov$ for small $h/L_0$ scale as $||\epsilon_p||_{L^2(\Omega)}(ov=75\%) = 0.499||\epsilon_p||_{L^2(\Omega)}(ov=0\%)$ and $||\epsilon_p||_{L^2(\Omega)}(ov=87.5\%) = 0.249 ||\epsilon_p||_{L^2(\Omega)}(ov=0\%)$, whose coefficients are expected to be $0.5$ and $0.25$ respectively as predicted by eq.~\eqref{eq:MeanVarOv}. 
This observation is intuitive as numerical differentiation amplifies the noise in a signal, while computing the derivatives of correlated noise is expected to amplify the noise less.
This test and the corresponding derivation given in appendix~\ref{sec:Appendix overlap:pressure reconstruction} indicates that high interrogation window overlap ratio reduces the error in the pressure calculation, at least by taming the noise amplification due to numerical differentiation.  

\section{A simple trick for more robust pressure calculation}
\label{sec:Trick}
The two major goals of the present work are to 
i) quantitatively explain and predict the optimal spatial resolution that minimizes the error in the calculated pressure field, and 
ii) quantify the minimum resolvable pressure computed from solving the PIV-based pressure Poisson equation. 
Our analysis shows that we may be able to avoid the concerns about the selection of the optimal spatial resolution, which has been verified and discussed in the literature \cite{charonko2010assessment,pan2016errorT,mcclure2017instantaneous,mcclure2017optimization}.

According to the analysis in eq.~\eqref{Aeq:E}, the major contributor for the error in the reconstructed pressure field is the squared terms in the data such as $(\partial \epsilon_u/\partial x)^2$ and $(\partial \epsilon_v/\partial y)^2$ when PIV error dominates the error propagation.
These terms are from the direct calculation of squared terms in eq.~\eqref{eq:PoissonEq2D}, which could be avoided for incompressible flows by leveraging a simple trick. 
The popular form of the pressure Poisson equation can be transformed by adding $(\nabla\cdot\bm{u})^2=0$ to the data. 
For example, in 2D, eq.~\eqref{eq:PoissonEq2D} becomes 
\begin{linenomath}
	\begin{equation}
		\label{eq:PoissonEq2DNew}
		\begin{split}
			\nabla^2 p &= - \left(\left(\frac{\partial u}{\partial x}\right)^2 + 2\frac{\partial u}{\partial y} \cdot \frac{\partial v}{\partial x} + \left(\frac{\partial v}{\partial y}\right)^2\right) + \left(\nabla \cdot \bm{u}\right)^2\\
			&=   2\left( \frac{\partial u}{\partial x} \frac{\partial v}{\partial y} -  \frac{\partial u}{\partial y} \frac{\partial v}{\partial x}\right)  \quad \text{in}  ~\Omega \in \mathbb{R}^2,
		\end{split}
	\end{equation}
\end{linenomath}
where $ (\nabla\cdot\bm{u})^2 = \frac{\partial u}{\partial x}^2 + 2\frac{\partial u}{\partial x} \frac{\partial v}{\partial y} + (\frac{\partial v}{\partial y})^2$. 
Eq.~\eqref{eq:PoissonEq2DNew} only contains cross terms. 
As long as the noise in different velocity components are not strongly correlated or biased, the numerically evaluated data do not necessarily suffer any bias. 
By avoiding the bias due to the squared terms, we may be able to reduce the $h^{-2}$ error terms in the calculated pressure field (e.g., in eq.~\eqref{eq:error} or eq.~\eqref{eq:error2}). 

\begin{figure}[!h]
	\centering
	\includegraphics[width=1\textwidth]{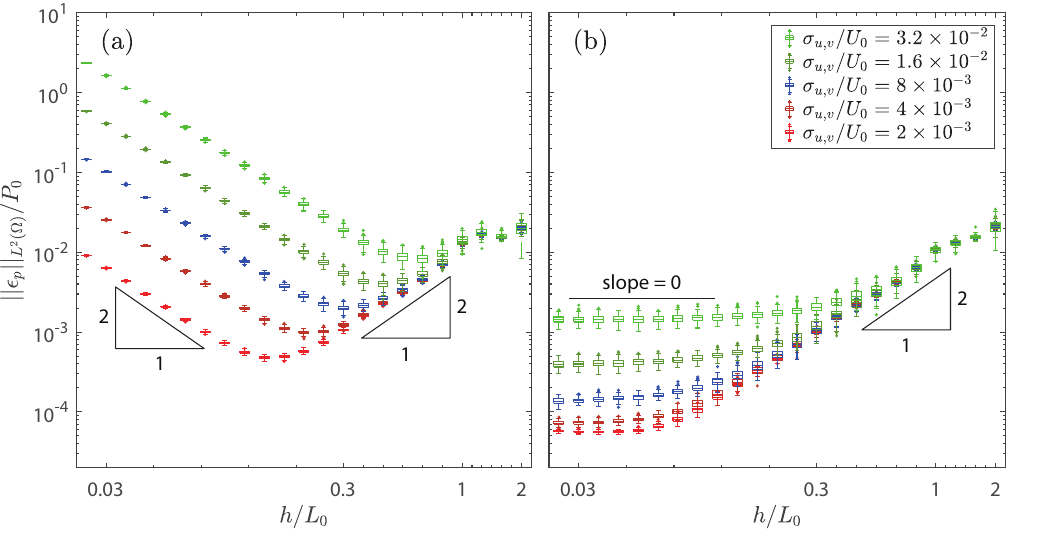}
	\caption{Error level of the calculated pressure field $||\epsilon||_{L^2(\Omega)}$ at various level of grid spacing $h$ with different point-wise Gaussian noise. 
	The pressure is calculated based on different formulation of data: (a)~eq.~\eqref{eq:PoissonEq2D} and (b)~~eq.~\eqref{eq:PoissonEq2DNew},  respectively. 
	The blue boxes represents the results based on the simulated data from \textit{Setup~1} above.
		\label{fig:Approach}}
\end{figure}

We use the same vortex flow as used in \textit{setup~1} to test this idea. 
The PIV data is generated by adding different point-wise independent Gaussian noise (i.e., $\epsilon_u \sim \mathcal{N}(0,\sigma_u^2)$, $\epsilon_v \sim \mathcal{N}(0,\sigma_v^2)$, $\sigma_u/U_0= \sigma_v/U_0 = 2\times10^{-3}, 4\times10^{-3}, 8\times10^{-3},1.6\times10^{-2},\text{~and~} 3.2\times10^{-2}$) to the uncontaminated velocity field. 
The synthetic flow with $\sigma_u/U_0= 8\times10^{-3}$ corresponds to \textit{setup~1} above.
We reconstruct the pressure field using eq.~\eqref{eq:PoissonEq2D}
and \eqref{eq:PoissonEq2DNew} on the contaminated PIV results and the error levels are presented in Fig.~\ref{fig:Approach}(a) and (b), respectively.

As shown in Fig.~\ref{fig:Approach}(a) the error level in the pressure field shows a V-shaped trend when eq.~\eqref{eq:PoissonEq2D} is used to compute the data.
The $-2$ slope when $h/L_0$ is small and the optimal spatial resolution has been discussed extensively above, so we don't focus on that here.
The caveat of the V-shaped error trend is that the use of eq.~\eqref{eq:PoissonEq2D} could make the pressure reconstruction too sensitive to the PIV error when the grid spacing is too small. 
When eq.~\eqref{eq:PoissonEq2DNew} is used to compute the data for the pressure Poisson equation, the trend of the error in the pressure field is L-shaped: when $h/L_0$ is small, $||\epsilon_{L^2(\Omega)}/P_0||$ plateaus as shown in Fig.~\ref{fig:Approach}(b). 
This is because eq.~\eqref{eq:PoissonEq2DNew} eliminates the bias due to the squared terms, and the cross terms do not contribute positive definite bias error to the data. 
Instead, the random error (and the high frequency components) from the cross terms are largely tamed by the low pass filter effect of solving the pressure Poisson equation \cite{de2012instantaneous,faiella2021error}. 
Thus, for the test cases with the same level of noise in the PIV data, the pressure field calculation based on eq.~\eqref{eq:PoissonEq2DNew} enjoys lower error especially when the spatial resolution is high (compare the boxes with the same color in Fig.~\ref{fig:Approach}(a) and (b)).
This simple trick using eq.~\eqref{eq:PoissonEq2DNew} can mitigate the risk of using too small $h/L_0$ that could cause excessive error in the pressure field if the popular formulation eq.~\eqref{eq:PoissonEq2D} is used.
When the spatial resolution is low, the error grows at a slope of $2$ when the spatial resolution reduces, which is the same as in Fig.~\ref{fig:Approach}(a).
This is expected as eq.~\eqref{eq:PoissonEq2DNew} only modifies the contribution from the PIV error and leaves the truncation error contribution unchanged.  
The intersection of the two trends (slope of zero for PIV-error dominated regime and the slope of two for numerical error dominated regime) is the critical resolution where increasing spatial resolution of the PIV data is neither helpful, nor harmful for pressure reconstruction. 
However determining the critical resolution is difficult, which may be affected by both the profile of the fluid flow and the profile of the error. 
We will leave this topic for future studies.

\section{Discussion and conclusions}
\label{sec:Conclusions}
We have provided a rigorous and general framework that decouples the contribution of numerical truncation and experimental noise to the pressure field reconstructed from PIV experiments. 
Based on this framework, we point out that the error propagation from the PIV-based velocity field measurements to the calculated pressure field is affected by many factors. 
The quality of the PIV experiments is only one aspect of many when quantifying the error that arises in the pressure field. 
Many other components of the problem are influential. For instance, the geometry and boundary conditions of the domain, the physical profile of the flow, and the numerical method (e.g., grid spacing and differencing scheme) of the pressure solver play a significant role. 
In this paper we have focused on one of these factors: how the spatial resolution of the velocity vector field from PIV impacts the error propagation. 

Previous independent studies with spatial resolutions that were too fine or too coarse resulted in significant errors in the calculated pressure field. 
We provide a precise theoretical estimation of the error level in the reconstructed pressure field, and the theory is validated by a large number of numerical experiments. 
Specifically, we give a precise description of the competition between the truncation error from the numerical schemes and the experimental noise from PIV experiments over the different spatial resolutions. 
When the spatial resolution is relatively fine, the error from the experimental data dominates the error propagation and when the spatial resolution is relatively coarse, truncation error due to the numerical scheme of the pressure solver governs the error propagation. 
Thus there may exist an optimal spatial resolution that minimizes the error propagation of a given flow. 
The corresponding minimum field-wide error level in the calculated pressure field can be considered the minimum resolvable pressure for the calculated field, or the effective sensitivity of the reconstructed pressure field. 
Since we find that the optimal spatial resolution is a rich function of the flow features, geometry of the flow domain, and the type of boundary conditions, as well as the quality of the PIV experiment, this means that PIV experiments to be used for pressure calculations must and can be carefully designed so that the optimal pressure estimation is achieved. 
Otherwise, large error could be introduced into the reconstructed pressure field. 

 In addition to the analytical results presented, we provide practical guidelines to estimate the error in the reconstructed pressure field. 
The coefficient of the truncation error $K_1$ and that of the PIV error $K_2$ are determined by Richardson extrapolation and adding artificial overwhelming noise, respectively. 
Numerical tests verify the effectiveness of this practical method we propose. 
We also analyzed and verified that high interrogation window overlap may reduce the error in the pressure calculation.
We also propose a simple trick that helps bypass the choice of optimal spatial resolution for incompressible flows: eliminating the squared terms in the data, which contribute a positive-definite error, can significantly reduce the error propagation.
Numerical tests show the advantage of this method, implying that the formulation of the solver matters.



We emphasize that the current research mainly focuses on a general framework that decouples error from the true value in the calculated pressure field.
The uncertainties in the calculated pressure field can then be directly analyzed. 
Although the framework in this work is general, the specific form of some of the pertinent equations (e.g., eq.~\eqref{eq:error1}) depends on the specific numerical schemes of the solver (e.g., second order central finite difference with structural grid spacing for the current consideration) and the error model (e.g., zero mean noise at each grid point) is not general. 
Different numerical schemes and more sophisticated models of the velocity field PIV-measure error will not fundamentally change the main results discussed above, and the approach taken here provides a guide for future investigations of such setups.

For a Poisson equation based approach to reconstruct the pressure field from PIV velocity data, the time derivative appears only through Neumann boundary conditions. 
Thus, the temporal resolution of the PIV data affects the error propagation mainly through Neumann boundaries, and it shows similar behavior of the effect of spatial resolution, as observed in \cite{charonko2010assessment,pan2016errorT}. 
A rigorous analysis (e.g., pursing a sharp estimate) of the effects of the boundary condition on the entire domain is beyond the scope of this paper and will be left for future research. 

\section*{Acknowledgment}
 We appreciate discussions with Jeff McClure and Dr. Serhiy Yarusevych.

\section*{Appendices}
\appendix
\numberwithin{equation}{section}

\section{Derivations of the error estimation}
\label{sec:Appendix derivation}
Consider a large domain in two dimensions (2D) with Dirichlet boundary conditions, which refers to the pressure Poisson equation in eq.~\eqref{eq:PoissonEq}. 
Using a five-point scheme on a structured mesh, a point-wise finite difference approximation of eq.~\eqref{eq:PoissonEq} is
\begin{linenomath}
	\begin{equation}
	\label{Aeq:NumericalPoisson}
 	\left.\nabla^2_h p\right|_{i,j} + \left.T_{\nabla^2 p}\right|_{i,j}+ \cdots \approx \left.f(\bm{u})\right|_{i,j} = - \left. \nabla \cdot \left( \left( \bm{u} \cdot \nabla\right) \bm{u} \right)\right|_{i,j} \quad \text{in} ~ \Omega, 
	\end{equation}
\end{linenomath}
where $\nabla^2_h$ denotes a numerical Laplacian with grid spacing $h$.
For example, evaluation of $\nabla^2_h$ at a grid point $(i,j)$ is 
\begin{linenomath}
	\begin{equation}
	\label{Aeq:NumericalLaplacian}
	\left.\nabla_h^2 p\right|_{i,j} = \frac{p_{i+1,j}+p_{i-1,j}+p_{i,j+1}+p_{i,j-1}-4p_{i,j}  }{h^2},
	\end{equation}
\end{linenomath}
and the corresponding leading order truncation error $\left.T_{\nabla^2 p}\right|_{i,j}$ is
\begin{linenomath}
	\begin{equation}
	\label{Aeq:Truncation}
	\left.T_{\nabla^2 p}\right|_{i,j} = - \frac{2}{4!}\left( \left. \frac{ \partial^4 p}{\partial x^4} \right|_{i,j}  +  \left. \frac{ \partial^4 p}{\partial y^4} \right|_{i,j}  \right) h^2. 
	\end{equation}
\end{linenomath}
This formulation is ignoring the effects of error in the velocity field. 
To retain such effects we recognize that the PIV velocity field ($\tilde{\bm{u}}$) contains error ($\bm{\epsilon}_u$), i.e. $\tilde{\bm{u}} = \bm{u} + \bm{\epsilon}_u$. 
This will lead to a reconstructed pressure field ($\tilde{p}$) contaminated by both the experimental noise and truncation numerical error, i.e. $\tilde{p} = p + \epsilon_p$, where, $\epsilon_p$ is the error in the calculated pressure field, and $p$ is the true value of the pressure field. 
Implemented numerically this is:
\begin{linenomath}
	\begin{equation}
	\label{Aeq:ErrorPoissonEq}
	\left.\nabla^2_h \tilde{p}\right|_{i,j} = \left.f(\tilde{\bm{u}})\right|_{i,j} \quad \text{in}~ \Omega. 
	\end{equation}
	Taking advantage of linearity of the Poisson operator, eq.~\eqref{Aeq:ErrorPoissonEq} becomes
	\begin{equation}
	\label{Aeq:ErrorPoissonEq2}
	\left.f(\tilde{\bm{u}})\right|_{i,j} = \left.\nabla^2_h p\right|_{i,j} + \left.\nabla^2_h \epsilon_{p}\right|_{i,j} \quad \text{in} ~ \Omega. 
	\end{equation}
\end{linenomath}

Comparing eq.~\eqref{Aeq:NumericalPoisson} and eq.~\eqref{Aeq:ErrorPoissonEq2}, we see that the numerically evaluated Laplacian of the calculated pressure error is:
\begin{linenomath}
	\begin{equation}
	\label{Aeq:Error}
	\left. \nabla^2_h\epsilon_{p}\right|_{i,j} = \left.T_{\nabla^2 p}\right|_{i,j} + \left.E_{\nabla^2 p}\right|_{i,j}  \quad \text{in} ~ \Omega, 
	\end{equation}
\end{linenomath}
where 
\begin{linenomath}
	\begin{equation}
	\label{Aeq:EL^2p}
	\left.E_{\nabla^2 p}\right|_{i,j} = \left.f(\bm{u})\right|_{i,j}  - \left.f(\tilde{\bm{u}})\right|_{i,j} =  \left. \nabla \cdot \left( {\bm{\epsilon}}_{u} \nabla \bm{u} + \bm{u}\nabla {\bm{\epsilon}}_{u}  + {\bm{\epsilon}}_{u} \nabla \bm{\epsilon}_u  \right) \right|_{i,j},
	\end{equation}
\end{linenomath}
is the error induced by the noisy PIV measurements. Eq.~\eqref{Aeq:Error} indicates that the total error in the reconstructed pressure field is influenced by two distinct factors: i) truncation error due to numerical schemes ($T_{\nabla^2 p}$) and ii) propagated errors from the velocity field due to noisy PIV experimental measurements ($E_{\nabla^2 p}$), an observation which is consistent with works in the area (e.g., \cite{charonko2010assessment,mcclure2017optimization,pan2016errorT}). 
More importantly, this formulation sets up a general framework that enables direct analysis of the contribution of each term. 

Now we decouple the contributions from $T_{\nabla^2 p}$ and $E_{\nabla^2 p}$ by first considering the scaling of each term with respect to the spatial resolution ($h$).
Recalling that a Poisson solver filters out the high frequency noises \citep{de2012instantaneous,pan2016error2,faiella2021error}, when the errors from PIV experiments are mainly high frequency random noise, rather than systematic biases, a major contribution from $E_{\nabla^2 p}$ would be the squared terms such as $(\partial \epsilon_u/\partial x)^2$ and $(\partial \epsilon_v/\partial y)^2$, which contribute a positive definite bias over the domain.
With this supposition we estimate $E_{\nabla^2 p}$ as
\begin{linenomath}
	\begin{equation}
	\label{Aeq:E}
	\left. E_{\nabla^2 p}\right|_{i,j} \approx  \left. \left(\frac{ \partial \epsilon_u}{\partial x} \right)^2 \right|_{i,j}  +  \left. \left(\frac{ \partial \epsilon_v}{\partial y} \right)^2 \right|_{i,j}. 
	\end{equation}
\end{linenomath}
Noting that in eq.~\eqref{Aeq:E}, the contribution of the $u$-component and the $v$-component are assumed to be decoupled and independent, the following derivation only takes the $u$-component as an example. If the velocity gradients are computed via a second order central difference scheme, e.g., $ \left.\frac{\partial \tilde{u}}{\partial x}\right|_{i,j} = \frac{\tilde{u}_{i+1,j} - \tilde{u}_{i-1,j}}{2h}$, then error in the velocity gradient fields is automatically computed in the same way:
\begin{linenomath}
	\begin{equation}
    \label{Aeq:PartialError}
	\left.\frac{\partial{\epsilon_u}}{\partial x}\right|_{i,j} = \frac{\epsilon_{u}|_{i+1,j} - \epsilon_{u}|_{i-1,j}}{2h}. 
	\end{equation}
\end{linenomath}


We assume that the noise in the velocity field at each nodal point is a zero mean random variable with a certain distribution $\mathcal{D}(0,\sigma_{u,v}^2),$  e.g., $\left. \epsilon_u \right|_{i,j} \sim \mathcal{D}(0,\sigma_{u}^2)$ and $\left. \epsilon_v \right|_{i,j} \sim \mathcal{D}(0,\sigma_{v}^2)$ for $u$ and $v$ components, respectively.
The expected value of the error in the velocity gradient in the $u$-component is:

\begin{linenomath}
	\begin{equation}
	\begin{aligned}
    \label{Aeq:MeanVar}
	\mathbb{E}\left[\left.\frac{\partial{\epsilon_u}}{\partial x}\right|_{i,j}\right] &= \mathbb{E}\left[\frac{\epsilon_{u}|_{i+1,j} - \epsilon_{u}|_{i-1,j}}{2h} \right] = \frac{\mathbb{E}\left[\epsilon_{u}|_{i+1,j} \right] - \mathbb{E}\left[\epsilon_{u}|_{i-1,j} \right]}{2h} = 0,
	\end{aligned}
	\end{equation}
\end{linenomath}
meaning that the error in the velocity gradient $\left.\frac{\partial{\epsilon_u}}{\partial x}\right|_{i,j}$ is also zero-mean. Thus, the variance of the error is 
\begin{linenomath}
	\begin{equation}
	\begin{aligned}
    \label{Aeq:MeanVar sqaur}
	\mathbb{E}\left[\left(\left.\frac{\partial{\epsilon_u}}{\partial x}\right|_{i,j}\right)^2\right] &= \mathbb{E}\left[\left(\frac{\epsilon_{u}|_{i+1,j} - \epsilon_{u}|_{i-1,j}}{2h} \right)^2 \right] \\
    &= \frac{\mathbb{E}\left[\epsilon^2_{u}|_{i+1,j} \right] + \mathbb{E}\left[\epsilon^2_{u}|_{i-1,j} \right] - 2\rho\left(\epsilon_{u}|_{i+1,j},  \epsilon_{u}|_{i-1,j}\right) \sqrt{\mathbb{E}\left[\epsilon^2_{u}|_{i+1,j} \right] \mathbb{E}\left[\epsilon^2_{u}|_{i-1,j}\right]}}{4h^2} \\
    &= \frac{1 - \rho\left(\epsilon_{u}|_{i+1,j},  \epsilon_{u}|_{i-1,j}\right)}{2h^2} \sigma_u^2. 
	\end{aligned}
	\end{equation}
\end{linenomath}
The variance of $\left.\frac{\partial{\epsilon_u}}{\partial x}\right|_{i,j}$ is a function of $\rho\left(\epsilon_{u}|_{i+1,j}, \epsilon_{u}|_{i-1,j}\right)$, which is the correlation coefficient of $\epsilon_{u}$ between two ``neighboring'' grid points (whose spacing is $2h$ due to the use of central finite difference scheme).\footnote{This equation applies when the second order central finite difference is used to calculate the source, as used through out the this paper. Other differential schemes may have different results.}  

For now, we assume the error in the velocity field at each mesh grid is a point-wise independent random variable\footnote{The goal of making this assumption is to streamline the structure of error estimation. 
Discussions about none-zero correlation cases can be found in Appendix~\ref{sec:Appendix overlap}}, thus,  $\rho\left(\epsilon_{u}|_{i,j},  \epsilon_{u}|_{k,l}\right) = 0$ and the squared error gradient at each grid point has a constant expectation, and eq.~\eqref{Aeq:MeanVar sqaur} reduces to
\begin{linenomath}
	\begin{equation}
	\begin{aligned}
    \label{Aeq:MeanVar sqaured rho=0}
\mathbb{E}\left[\left(\left.\frac{\partial{\epsilon_u}}{\partial x}\right|_{i,j}\right)^2\right] = \frac{\sigma_u^2}{2h^2}. 
	\end{aligned}
	\end{equation}
\end{linenomath}

Similarly, the same deviation can be applied to the gradient of velocity error in the $v$-component, which takes the same form, i.e., 
\begin{linenomath}
	\begin{equation}
	\begin{aligned}
    \label{Aeq:MeanVar sqaured rho=0 v}
\mathbb{E}\left[\left(\left.\frac{\partial{\epsilon_v}}{\partial y}\right|_{i,j}\right)^2\right] = \frac{\sigma_v^2}{2h^2}. 
	\end{aligned}
	\end{equation}
\end{linenomath}
The rub of the matter is that the
error introduced by the noise from the experiments will scale as
\begin{linenomath}
	\begin{equation*}
	E_{\nabla^2p} \sim O(h^{-2}).
	\end{equation*}
\end{linenomath}
Compared to the contribution from truncation error (see eq.~\eqref{Aeq:Truncation}):
\begin{linenomath}
	\begin{equation*}
	T_{\nabla^2p} \sim O(h^{2}),
	\end{equation*}
\end{linenomath}
it is clear that when $h$ is small, the experimental error dominates the error propagation, and the contribution from truncation error vanishes. Similarly, when $h$ is large, the numerical truncation error is dominant, but the impact from experimental error is negligible. Each of these terms is analyzed in more detail below.

For the truncation error, rewriting eq.~\eqref{Aeq:Truncation} leads to a point-wise description over the domain: \begin{linenomath}
	\begin{equation}
	\label{Aeq:Truncation2}
	\left. T_{\nabla^2 p} \right|_{i,j} = - \frac{2}{4!}\left( \left. \frac{ \partial^4 p}{\partial x^4} \right|_{i,j}  +  2\left. \frac{ \partial^4 p}{\partial x^2 \partial y^2} \right|_{i,j} + \left. \frac{ \partial^4 p}{\partial y^4} \right|_{i,j}  \right)h^2  +  2\frac{2}{4!}\left. \frac{ \partial^4 p}{\partial x^2 \partial y^2} \right|_{i,j}h^2,  
	\end{equation}
\end{linenomath}
and integrating twice we have the corresponding truncation error of the pressure field: 
\begin{linenomath}
	\begin{equation}
	\label{Aeq:Truncation3}
	\epsilon_{p,T} = \frac{1}{12}\left( -\nabla^2 p + 2\nabla^{-2}\frac{\partial^{4} p}{\partial^2x \partial^2y} \right)h^2,
	\end{equation}
\end{linenomath}
where $\nabla^{-2}$ is the inverse Laplacian which is specifically dependent on the domain and type of boundary conditions. Thus, the total error introduced by the truncation error can be estimated as 
\begin{linenomath}
	\begin{equation}
	\label{Aeq:Truncation4}
	\| \epsilon_{p,T} \| \lesssim C_1 \left( \left\Vert \frac{\partial^2 p}{\partial x^2} \right\Vert_{L^2(\Omega)}+ 2\left\Vert \nabla^{-2}\frac{\partial^{4} p}{\partial^2 x \partial^2 y} \right\Vert_{L^2(\Omega)} +   \left\Vert \frac{\partial^2 p}{\partial y^2} \right\Vert_{L^2(\Omega)}\right)h^2, 
	\end{equation}
\end{linenomath}
where $C_1=1/12$ is a constant inherited from the Taylor expansion relevant to the specific numerical solution of the Poisson equation.

The error arising in the experimental velocity field when $h \rightarrow 0$ is represented by point-wise squared error gradients implying that $E_{\nabla^2 p}$ can be split into high frequency random components and a uniform non-zero bias.
The error introduced into the data field can be estimated as
\begin{linenomath}
	\begin{equation}
	\label{Aeq:EL^2p2}
	E_{\nabla^2p} \approx \frac{1}{2h^2} \left( \sigma_u^2 + \sigma_v^2 \right),
	\end{equation}
\end{linenomath}
With the approach developed in \cite{pan2016error1,faiella2021error} we can bound the error in the pressure field due to experimental error in the velocity field as
\begin{linenomath}
	\begin{equation}
	\label{Aeq:E_exp}
	||\epsilon_{p,E}||_{L^2(\Omega)}  \lesssim C_0 C_2 \left(\frac{\sigma_u^2 + \sigma_v^2}{2} \right) h^{-2}. 
	\end{equation}
\end{linenomath}
$C_2$ can be considered as the amplification of the error level in the pressure field ($\epsilon_p$) to the error level in the data ($\epsilon_f$) when the data $f$ has the `worst' profile. 
In other words, $C_2 = \frac{ ||\epsilon_{p}||_{L^2(\Omega)}}{||\epsilon_f||_{L^2(\Omega)}}$. 
Assuming a large square $L \times L$ domain with Dirichlet boundary conditions, the optimal Poincar\'e constant is given by $C_2= \frac{L^2}{2\pi^2}$. 
$C_0$ measures the difference between a uniform error in the data and the ``worst'' possible error field. 
For the 2D example with Dirichlet boundary conditions as considered here, this ratio is the square of the 1D case (see \cite{faiella2021error} for greater details), thus $C_0 \approx 0.901^2$. 

Combining eq.~\eqref{Aeq:Truncation4} and eq.~\eqref{Aeq:E_exp}, we have an estimate of the total error in the reconstructed pressure field: 
\begin{linenomath}
	\begin{equation}
	\begin{split}
	\label{Aeq:Error_total}
	||\epsilon_{p}||_{L^2(\Omega)} 	&= ||\epsilon_{p,T} + \epsilon_{p,E}||_{L^2(\Omega)} \\ &\lesssim ||\epsilon_{p,T} ||_{L^2(\Omega)} + ||\epsilon_{p,E}||_{L^2(\Omega)} \\ & \approx  C_1 \left( \left\Vert \frac{\partial^2 p}{\partial x^2} \right\Vert_{L^2(\Omega)}+ 2\left\Vert \nabla^{-2}\frac{\partial^{4} p}{\partial^2 x \partial^2 y} \right\Vert_{L^2(\Omega)} +   \left\Vert \frac{\partial^2 p}{\partial y^2} \right\Vert_{L^2(\Omega)}\right)h^2 +  C_0C_2 \left(\frac{\sigma_u^2 + \sigma_v^2}{2} \right) h^{-2},
	\end{split}
	\end{equation}
\end{linenomath}
which is eq.~\eqref{eq:error1} in the body of the paper. 
The derivation of the error estimation provided here is for a simplified setting, but other cases (e.g., with different boundary conditions, dimensions or numerical schemes) could be determined in a similar manner.

\section{Effect of the interrogation overlap ratio and correlated error}
\label{sec:Appendix overlap}

Employing overlapped interrogation windows is common practice in PIV, however,  a non-zero overlap ratio causes correlated measurement and noise \citep{Wieneke2017PIV,howell2018distribution}.
In this section, we discuss how the interrogation window overlap ratio, and the corresponding correlation of the error in the PIV results, may affect the error propagation from the velocity field to the constructed pressure field.

Currently, there is no concrete theory that  models/explains how the overlap ratio of the interrogation window affects the error correlation.
However, some observations showed that the random error of real PIV results have a specific character: the covariance with respect to the distance to nearby vectors show a triangular or bell curve-like profile \citep{Wieneke2017PIV,howell2018distribution} and spatial filters \citep{Wieneke2017PIV} and Cholesky decomposition \citep{mcclure2017instantaneous} are proposed to generate such results. 
We next model the correlated random noise in  PIV using a ``top-hat'' kernel as proposed in  \cite{Wieneke2017PIV}.

\subsection{Modeling correlated random noise in PIV}
\label{sec:Appendix overlap:modeling}

Consider independent zero mean Gaussian variables $X_{i,j}$ on an $M \times N$ structured mesh ($i = 1, 2, ... , M$ and $j = 1, 2, ... , N$) where each $X_{i,j}$ is a point-wise independent zero mean variable: $X_{i,j} \sim \mathcal{N}(0,\sigma^2)$. We apply a moving average filter (the kernel is $\frac{1}{4}\bm{J}_{2\times2}$, where $\bm{J}$ is an all-ones matrix) to $X_{i,j}$ throughout the domain. At each center of the window, the filtered variable is 
\begin{equation}
Y_{i,j} = \frac{1}{4}\sum_{i-1}^{i+1}  \sum_{j-1}^{j+1}  X_{i,j},
\label{Aeq:kernel}
\end{equation}
and $Y_{i,j}$ is also a Gaussian variable:
\begin{equation*}
Y_{i,j} \sim \mathcal{N}\left(0,\frac{\sigma^2}{4}\right).
\label{eq:sigma_X}
\end{equation*}
Note that $Y_{i,j}$ of course enjoys a smaller variance than $X_{i,j}$ owing to the filtering.

\begin{figure}[!h]
	\centering
	\includegraphics[width=0.9\textwidth]{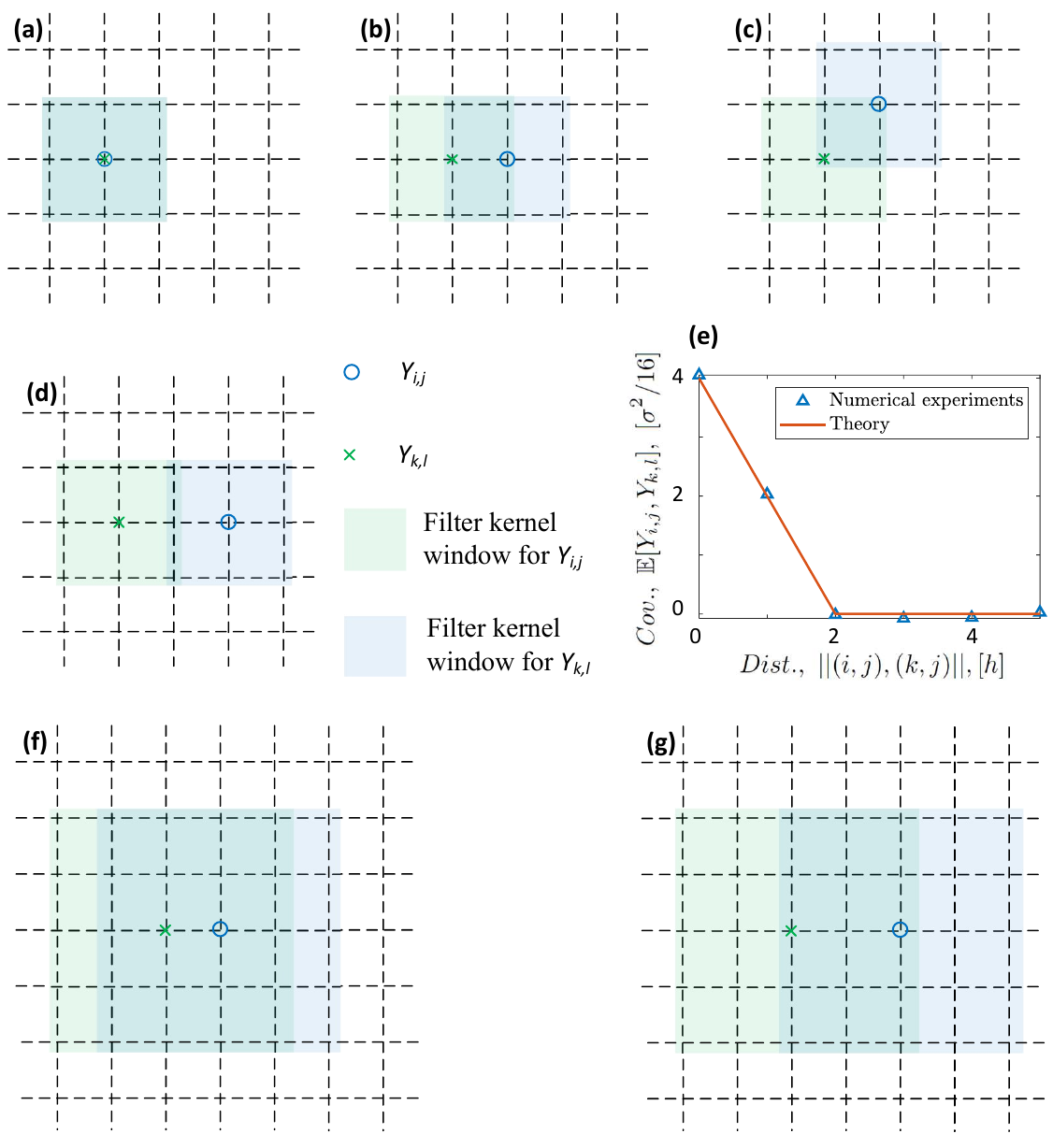}
	\caption{(a-d) The overlapped area of the neighboring $\bm{J}_{2\times2}$ kernels, which refers to $ov=50\%$. The distances of the kernels $||(i,j)-(k,l)||$ are $0$, $h$, $\sqrt{2}h$, $2h$, respectively. The green and blue boxes are the neighboring kernels. The central crosses and circles represent the filtered `data' by each kernel. The intersection between green area and blue area represents the magnitude of covariance. (e) A comparison of eq.~\eqref{eq:Covar} against a numerical test. The blue triangles are the the calculation results of the numerical test. The orange curve is the prediction given by eq.~\eqref{eq:Covar}. (f-g) The representative overlapped area of the neighboring $\bm{J}_{4\times4}$ kernels, which refers to $ov=75\%$. The distances of the kernels $||(i,j)-(k,l)||$ are $h$, $2h$, respectively.}
	\label{fig:Correlation_Err}
\end{figure}

We now calculate covariance of the spatially filtered variable $Y_{i,j}$ and its neighbor $Y_{k,l}$. The value of the covariance $\mathbb{E}[Y_{i,j}Y_{k,l}]$ depends on the distance ($||(i,j),(k,l)||$) between the two points in the domain, which determines the overlapped area of the neighboring filter kernel windows (see Fig.~\ref{fig:Correlation_Err}(a-d) for an illustration): 
\begin{equation}
\mathbb{E}[Y_{i,j}Y_{k,l}] = \frac{\sigma^2}{16} \begin{cases}
4, & ||(i,j),(k,l)|| = 0 \\
2, & ||(i,j),(k,l)|| = h \\
1, & ||(i,j),(k,l)|| = \sqrt{2}h \\
0, & ||(i,j),(k,l)|| \ge 2h. \\
\end{cases} \quad
\label{eq:Covar}
\end{equation}

The orthogonal cases of eq.~\eqref{eq:Covar}, corresponding to the use of central finite difference schemes in the PIV-based pressure calculation (e.g., eq.~\eqref{Aeq:PartialError}, are illustrated  in Fig.~\ref{fig:Correlation_Err}(b \& d). 
Fig.~\ref{fig:Correlation_Err}(e) shows comparison of eq.~\eqref{eq:Covar} (red curve) against statistics from a numerical test (blue triangles) similar to the ``top-hat'' filtering used in \cite{Wieneke2017PIV}. 
The covariance linearly decays when the distance of the two variables increases and leads to a triangular covariance profile.
Particularly, $Y_{i,j}$ is a zero mean Gaussian noise field with non-zero correlation, which is a model of correlated PIV error from processing with a 50\% overlap interrogation window. Similar properties and experimental tests can be found in \cite{Wieneke2017PIV,howell2018distribution}.

To evaluate the error propagation of the correlated noises based on the finite difference Poisson solver in this work, we start with calculating the statistics of $dY_{i,j}=Y_{i+1,j} - Y_{i-1,j}.$ 
It is apparent that the expected value $\mathbb{E}[dY_{i,j}] = 0$,  since $\mathbb{E}[Y_{i,j}]=0$. 
The variance can be calculated as 
 $$\mathbb{E}[dY_{i,j}^2] = \mathbb{E}[(Y_{i+1,j} - Y_{i-1,j})^2]= 2\mathbb{E}(Y_{i,j}^2) - 2\mathbb{E}[Y_{i+1,j} Y_{i-1,j}].$$  
 Noting that $||(i+1,j),(i-1,j)|| = 2h$ and applying eq.~\eqref{eq:Covar}, the above equation leads to 
 \begin{equation}
\mathbb{E}[dY_{i,j}^2] = \frac{1}{2}\sigma^2.
\label{eq:sigma_Y}
 \end{equation}
Combining eq.~\eqref{eq:sigma_X} and eq.~\eqref{eq:sigma_Y}, the variance of $dY_{i,j}$ is exactly twice of the variance of $Y_{i,j}$: 
 \begin{equation}
\mathbb{E}[dY_{i,j}^2] = 2\mathbb{E}[Y_{i,j}^2].
\label{eq:sigma_ratio}
 \end{equation}

\subsection{In the context of PIV-based pressure reconstruction}
\label{sec:Appendix overlap:pressure reconstruction}
The dominant contributor in the source of the pressure Poisson equation are the squared terms such as
 \begin{equation}
\left.\left(\frac{\partial \epsilon_u}{\partial x}\right)^2\right|_{i,j} = \left(\frac{\epsilon_{u_{i+1,j}} - \epsilon_{u_{i-1,j}}}{2h}\right)^2.
\label{eq:squared_difference}
 \end{equation}
We assume that $\epsilon_{u_{i,j}}$ has point-wise variance $\sigma^2_u$, and linearly decaying correlation with neighboring cells no further than two grid spacings ($2h$) apart. It is important to note that the variance $\sigma^2_u$ introduced here is the PIV error after window overlap, which corresponds to $\mathbb{E}[Y_{i,j}^2]$ in eq.~\eqref{eq:sigma_ratio}. 
Using a central finite difference scheme, the expectation of $\left(\frac{\partial \epsilon_u}{\partial x}\right)^2$ is
\begin{equation}
\mathbb{E}\left[\left(\frac{\partial \epsilon_u}{\partial x}\right)^2\right]_{ov=50\%} = \frac{\sigma_u^2}{2h^2},
\label{eq:error_source50}
\end{equation}  
which corresponds to PIV data generated on a 50\% overlap ratio, and the same as the `unfiltered' error field.
In fact, for the 50\% overlap interpretation window, $\epsilon_{u_{i+1,j}}$ and $\epsilon_{u_{i-1,j}}$ are uncorrelated.

The above analysis (from $\eqref{Aeq:kernel}$ to $\eqref{eq:sigma_ratio}$) gives an example for calculating the expected error for velocity gradients with $ov=50\%$. 
Noting that the filter kernel for generating PIV data is equivalent to the interrogation window in PIV, the analysis can be extended to a general calculation procedure for any interrogation window overlap ratio. 

Varying overlap ratio $ov$ connects the grid spacing ($h$) of the PIV data and the size of interrogation window ($Sz$):
\begin{linenomath}
  \begin{equation}
  \label{Aeq:Sz}
        Sz = \frac{h}{1-ov},
  \end{equation}
\end{linenomath}
where $Sz$ is also equivalent to that of the space average kernel.
For a given grid spacing $h$, $Sz$ can be calculated for each $ov$ according to eq.~\eqref{Aeq:Sz}, and similar to eq.~\eqref{Aeq:kernel}, a filter kernel of $\frac{1}{Sz\times Sz}\bm{J}_{Sz\times Sz}$ can be constructed for each corresponding $ov$. 
As central finite difference schemes (i.e.,  eq.~\eqref{Aeq:PartialError}) are applied to compute the derivatives, only vectors with distance  $||(i,j),(k,l)|| = 2h$ in eq.~\eqref{eq:Covar} affect the computation.
When $||(i-1,j),(i+1,j)|| = 2h \ge Sz$, which is corresponding to $ov\le50\%$ (see eq.~\eqref{Aeq:Sz}), there is no overlapping area between the two kernels, and the vectors at the two grid points ($\epsilon_{u_{i-1,j}}$ and $\epsilon_{u_{i+1,j}}$) are independent and the covariance $\mathbb{E}[Y_{i+1,j},Y_{i-1,j}]$ is zero.
When $50\% < ov < 100\%$, the covariance is no longer zero and the correlation coefficient can be calculated by evaluating the ratio of the overlapped area to the area of the kernel, i.e.,
\begin{linenomath}
	\begin{equation}
    \label{Aeq:CorrCoefSz}
        \rho\left(\epsilon_{u}|_{i-1,j},  \epsilon_{u}|_{i+1,j}\right) = \dfrac{\left(Sz-||(i-1,j), (i+1,j)||\right)\times Sz}{Sz\times Sz},
	\end{equation}
\end{linenomath}
where $\rho\left(\epsilon_{u}|_{i-1,j},  \epsilon_{u}|_{i+1,j}\right)$ is the correlation coefficient of $\epsilon_{u}|_{i-1,j}$ and $\epsilon_{u}|_{i+1,j}$. 
The numerator $\left(Sz-||(i-1,j), (i+1,j)||\right)\times Sz$ represents the overlapped area (see Fig.~\ref{fig:Correlation_Err}(b,~d,~f,~and~g) for illustration) and the denominator $Sz\times Sz$ is the area of each kernel.
Considering $||(i-1,j), (i+1,j)|| = 2h$ due to the use of central finite difference schemes and recalling eq.~\eqref{Aeq:Sz}, the correlation coefficient can be written as
\begin{equation}
\rho\left(\epsilon_{u}|_{i-1,j},  \epsilon_{u}|_{i+1,j}\right)  = 
\begin{cases}
0, & ov \le 50\% \\
2(1-ov), &  50\%< ov < 100\%.  
\end{cases} \quad
\label{eq:error_source_G}
\end{equation}  
Then eq.~\eqref{Aeq:MeanVar sqaur} can be written as:
\begin{linenomath}
	\begin{equation}
	\begin{aligned}
    \label{Aeq:MeanVarOv}
	\mathbb{E}\left[\left(\left.\frac{\partial{\epsilon_u}}{\partial x}\right|_{i,j}\right)^2\right] =
    \begin{cases}
         \dfrac{\sigma_u^2 }{2h^2} , & ov \leq 50\% \\
        \dfrac{(1 - ov)\sigma_u^2}{h^2}, & 50\%<ov\leq100\%
    \end{cases}
	\end{aligned}
	\end{equation}
\end{linenomath}
Therefore, eq.~\eqref{Aeq:MeanVarOv} gives a framework to calculate the effect of overlap ratio.

For example, the calculation of 75\% overlap interrogation window corresponds to a filter kernel of $\frac{1}{16}\bm{J}_{4\times 4}$ (shown in Fig.~\ref{fig:Correlation_Err}(f \& g)), and leads to 
\begin{equation}
\mathbb{E}\left[\left(\frac{\partial \epsilon_u}{\partial x}\right)^2\right]_{ov=75\%}  = \frac{\sigma_u^2}{4h^2},
\label{eq:error_source75}
\end{equation}  
which contributes half as much error as uncorrelated noise (compare eq.~\eqref{eq:error_source50} and eq.~\eqref{eq:error_source75}). 

Some typical cases of eq.~\eqref{Aeq:CorrCoefSz}, that correspond to `data' from central finite difference schemes and their corresponding filter kernel  used in the PIV-based pressure calculation, are shown in Fig.~\ref{fig:Correlation_Err}(d \& g) for $50\%$ and $75\%$ cases, respectively. 
Fig.~\ref{fig:Correlation_Err} shows comparison of eq.~\eqref{Aeq:MeanVarOv} against statistics from a numerical test, where a moving average filter is used to simulate the smear effect caused by an overlapped interrogation window. 
The covariance linearly decays when the distance of two variables increases and leads to a triangle covariance profile. 
Similar properties and experimental tests can be found in \cite{Wieneke2017PIV,howell2018distribution}.

The effect of the interrogation window overlap in eq.~\eqref{Aeq:MeanVarOv} is fundamentally the same for the uncertainty quantification for the vorticity calculation in \cite{sciacchitano2016piv}:
\begin{equation}
U_{\omega} = \frac{U}{d}\sqrt{1-\rho(2d)},
\label{eq: vortixity UQ}
\end{equation}  
where $U$ is the uncertainty of velocity, $d$ is the grid spacing, and $\rho(2d)$ is the correlation coefficient between the grid points whose distance is $2d$. 
The difference between eq.~\eqref{eq: vortixity UQ} and eq.~\eqref{Aeq:MeanVarOv} is that the computation of eq.~\eqref{Aeq:MeanVarOv} is quadratic, whereas  eq.~\eqref{eq: vortixity UQ} arises from the linear combination of velocity gradients.

\section{Remark on domain size and characteristic length scales}
\label{sec:Appendix domain size}
In this work, the choice of the characteristic length (or the reference length scale) may be arbitrary, and does not appear in the final error estimates. However, in practice, we recommend paying close attention to two important length scales: i) the smallest length scale of interest in the flow, and ii) the length scale of the dominant flow structure. As one considers these length scales two important rules of thumb should be remembered.

There is a danger in selecting a characteristic length scale that is much smaller than the given domain as this could lead to relatively large error (from the large domain) when compared to the pressure changes generated by the small-scale flow structures. In such cases, if the pressure change (\textit{true value}) induced by the small-scale flow structures of interest is comparable to the average \textit{error} in the pressure field over the domain (the non-dimensional error level is close to unity), the current pressure reconstruction setup (PIV experimental set up, PIV resolution choice, pressure solver choice, etc.) cannot resolve the pressure field corresponding to this small-scale flow. {Thus, for a large domain, the dominant flow structure (usually a larger scale than the small scale structures) should be used as the dimensional scale.}

If the small-scale flow structures and corresponding pressure field must be resolved, the obvious thing to do is optimize the pressure reconstruction set up (e.g., reducing the error in the PIV experiments, adjusting the boundary conditions, optimizing the pressure field, etc.). Another direct solution is to {shrink the domain size to a scale that is comparable to the small-scale flow structures.} With this adjustment, the non-dimensional domain size will be close to unity, and the small-scale structures will become the dominant features in the domain and the error levels will be smaller than the structures of interest.


\bibliographystyle{aps-nameyear} 
\bibliography{PivPressureLibrary}
\end{document}